\documentclass[oldversion]{aa}  
\usepackage{graphicx,amsmath,subfigure}
\usepackage{amssymb}
\usepackage{color}
\usepackage{natbib}		%for A&A : reimplements the LaTeX \cite command 
\usepackage{url}
\usepackage{multirow}
\bibpunct{(}{)}{;}{a}{}{,} % to follow the A&A style

\def\approxinf{%
  \def\p{%
    \setbox0=\vbox{\hbox{$<$}}%
    \ht0=0.6ex \box0 }%
  \def\s{%
    \vbox{\hbox{$\sim$}}%
  }%
  \mathrel{\raisebox{0.7ex}{%
      \mbox{$\underset{\s}{\p}$}%
    }}%
}

\begin{document}

   \title{Radiative braking in the extended exosphere of GJ\,436 b}
   				   
   \author{
   V.~Bourrier\inst{1},
   D.~Ehrenreich\inst{1},\and
   A.~Lecavelier des Etangs\inst{2}}
   
\authorrunning{V.~Bourrier et al.}
\titlerunning{Radiative braking in GJ\,436 b exosphere}

\offprints{V.B. (\email{vincent.bourrier@unige.ch})}

\institute{
Observatoire de l'Universit\'e de Gen\`eve, 51 chemin des Maillettes, 1290 Sauverny, Switzerland\and 
Institut d'astrophysique de Paris, UMR7095 CNRS, Universit\'e Pierre \& Marie Curie, 98bis boulevard Arago, 75014 Paris, France 
}
   
   \date{} %Received ...; accepted ...}
 
  \abstract
{The recent detection of a giant exosphere surrounding the warm Neptune GJ\,436 b has shed new light on the evaporation of close-in planets, revealing that moderately irradiated, low-mass exoplanets could make exceptional targets for studying this mechanism and its impact on the exoplanet population. Three HST/STIS observations were performed in the Lyman-$\alpha$ line of GJ\,436 at different epochs, showing repeatable transits with large depths and extended durations. Here, we study the role played by stellar radiation pressure on the structure of the exosphere and its transmission spectrum. We found that the neutral hydrogen atoms in the exosphere of GJ\,436 b are not swept away by radiation pressure as shown to be the case for evaporating hot Jupiters. Instead, the low radiation pressure from the M-dwarf host star only brakes the gravitational fall of the escaping hydrogen toward the star and allows its dispersion within a large volume around the planet, yielding radial velocities up to about -120\,km\,s\,$^{-1}$ that match the observations. We performed numerical simulations with the EVaporating Exoplanets code (EVE) to study the influence of the escape rate, the planetary wind velocity, and the stellar photoionization. While these parameters are instrumental in shaping the exosphere and yield simulation results in general agreement with the observations, the spectra observed at the different epochs show specific, time-variable features that require additional physics.}


   \maketitle

\section{Introduction}
\label{intro} 

\subsection{Atmospheric escape}

More than 40\% of the known exoplanets orbit extremely close to their star ($\approxinf$0.1\,au), raising many questions about the state of their atmosphere and its interactions with the host star. Theoretical studies indicate that the intense stellar X-ray and extreme ultraviolet energy input into the lower thermosphere of hydrogen-rich planets can lead to a hydrodynamic expansion of the upper gas layers. If the flow of expanding hydrogen reaches high velocities or high altitudes, it can escape the gravitational pull of the planet in large amounts, leading to the ``evaporation" or ``blow-off" of the atmosphere (\citealt{Lammer2003}; \citealt{Lecav2004}; \citealt{Yelle2006}; \citealt{GarciaMunoz2007}; \citealt{Koskinen2007}; \citealt{MurrayClay2009}; \citealt{Koskinen2013a,Koskinen2013b}). \\
Transit spectroscopy in the ultraviolet has proved an invaluable tool for studying the outer regions of an evaporating planet. Because of its expansion, the atmosphere produces a much deeper absorption than the planetary disk when observed in the lines of elements that are abundant at high altitudes and/or associated with strong electronic transitions. Observations in the bright stellar Lyman-$\alpha$ line of neutral hydrogen have led to the detections of evaporation from the hot-Jupiter HD\,209458b (\citealt{VM2003}), HD\,189733b (\citealt{Lecav2010}; \citealt{Lecav2012}; \citealt{Bourrier2013}), the hot-Neptune GJ\,436b (\citealt{Kulow2014}; \citealt{Ehrenreich2015}), and the warm-Jupiter 55 Cnc b (\citealt{Ehrenreich2012}). Heavier metals and ions can be carried upward through collisions with the expanding flow of hydrogen, which was confirmed by detection of several such species at high altitudes around HD\,209458 b (O$^{0}$, C$^{+}$, \citealt{VM2004}, \citealt{Linsky2010}; Mg$^{0}$, \citealt{VM2013}; and tentatively Si$^{2+}$, \citealt{Linsky2010}; \citealt{Ballester2015}) and HD\,189733 b (O$^{0}$ and possibly C$^{+}$, \citealt{BJ_ballester2013}), and allowed detection of an extended exosphere around the hot-Jupiter WASP-12 b through transit observations in the near-UV (Mg$^{+}$, Fe$^{+}$; \citealt{Fossati2010}, \citealt{Haswell2012}).\\
Observations at high-resolution in the UV have also been used to probe the structure of these extended exospheres, revealing that they are shaped by interactions with the host star, such as radiation pressure, stellar wind, and magnetic interactions (e.g., \citealt{Llama2011}, \citealt{Bourrier_lecav2013}; \citealt{Kislyakova2014}). Observations at different epochs in the Lyman-$\alpha$ line of HD\,189733 b led further to the first detection of temporal variations in the physical properties of an extended exosphere, which probably come from the energetic and/or dynamical influence of the host star (\citealt{Lecav2012}; \citealt{Bourrier2013}). \\
%Small mass planets
Many close-in exoplanets are expected to show signs of evaporation. The impact of this phenomenon on the formation and the evolution of planets has been studied through energy diagrams (\citealt{Lecav2007}, \citealt{Ehrenreich_desert2011}, \citealt{Lammer2009a}) and theoretical modeling applied to specific systems (e.g., \citealt{Lopez2012,Lopez2013}) or populations of exoplanets (e.g., \citealt{Kurokawa2014}). While massive gaseous giants like HD\,209458b or HD\,189733 b are subjected to moderate escape rate, leading to the loss of a few percentage points of the planet mass over the lifetime of the system, low-density planets like mini-Neptunes or super-Earths with a large volatile envelope may be significantly eroded by evaporation. This scenario is supported by trends in the exoplanet population: the lack of sub-Jupiter size planets (\citealt{Lecav2007}; \citealt{Davis2009}, \citealt{beauge2013}) and the large number of close-in super-earths (\citealt{howard2012}) are consistent with the formation of rocky planetary remnants through the evaporation of their hydrogen/helium envelope. It is thus crucial to observe the atmosphere of close-in Neptune-mass planets to understand in which conditions their evaporation can lead to the formation of atmosphereless remnant cores like CoRoT-7b and Kepler-10 b. 

\subsection{The case of GJ\,436 b}

In this framework, the warm Neptune GJ\,436 b (\citealt{Butler2004}; \citealt{Gillon2007}) is a good target for transit observations in the Lyman-$\alpha$ line, as it orbits a bright star (V=10.7) whose close proximity to the Earth (d=10.3\,pc) limits the effects of interstellar medium absorption in the core of this line (\citealt{Ehrenreich2011}). The low mass of GJ\,436 b ($M_{p}$=23.1\,$M_{Earth}$), the moderate irradiation from its M-dwarf host star ($a$=0.029\,au; \citealt{Ehrenreich2015}), and the puzzling eccentricity of its orbit ($e$=0.16, \citealt{Beust2012} and references therein) offer the opportunity to study a new evaporation regime. \citet{Kulow2014} reported the detection of a deep absorption signature in the stellar Lyman-$\alpha$ line starting near the end of the optical transit, thought to be caused by the occultation from a large cometary tail of neutral hydrogen H$^{0}$. However, interpretation of these observations was misled by the use of an inaccurate transit ephemeris and by the lack of an out-of-transit reference for the flux in the stellar Lyman-$\alpha$ line (\citealt{Ehrenreich2015}). Using two additional HST observations, \citet{Ehrenreich2015} revealed a much deeper H\,{\sc i} absorption that starts several hours before the optical transit ($\sim$2 - 4\,h) and that may remain detectable for more than 20\,hours afterward. GJ\,436 b is thus not only trailed by a cometary tail, but also surrounded by a giant coma of neutral hydrogen extending tens of planetary radii in front of the planet.\\
Our goal is to investigate the effects of radiation pressure and the basic physical parameters of an evaporating exoplanet on the structure of GJ\,436 b hydrogen exosphere and its transmission spectrum. To this aim we reconstructed the intrinsic Lyman-$\alpha$ line of the host star and compared the results of 3D numerical simulations of extended atmospheres based on the EVaporating Exoplanets code (EVE) with the three sets of transit observations. Observations are presented in Sect.~\ref{obs}, and the reconstruction of the intrinsic stellar Lyman-$\alpha$ line is explained in Sect.~\ref{lalpha}. The EVE code is described in Sect.~\ref{model} and used in Sect.~\ref{struct} to study the properties of GJ\,436 b exosphere in the low radiation pressure field of its host star. In Sect.~\ref{constraints} we discuss the part played by radiation pressure in the observed features and investigate several mechanisms that could also intervene. We summarize our conclusions in Sect.~\ref{conclu}.\\

%%%%%%%%%%%%%%%%%%%%%%%%%%%%%%%%%%%%%%%%%%%%%%%%%%%%%%%%%%%%%%%%%%%%%%%%%%%%%%%%%%%%%%%%%%%
\section{HST/STIS observations in the Lyman-$\alpha$ line}
\label{obs}

We used the three existing transit data sets of GJ\,436b observations in the H\,{\sc i} Lyman-$\alpha$ line (1215.6702\,\AA), all taken with similar settings with the Space Telescope Imaging Spectrograph (STIS) instrument onboard the Hubble Space Telescope (HST). The HST data consist of time-tagged spectra obtained with the G140M grating, with a spectral resolution of $\sim$20\,km\,s$^{-1}$ at 1215.67\,\AA. Data was reduced as described in \citealt{Ehrenreich2015}. Throughout the paper we use the term transit to refer to the transit of the extended exosphere of GJ\,436 b, while we explicitely speak of optical transit to refer to the occultation caused by the planetary disk alone. Each data set was obtained through four HST orbits performed at different phases near the optical transit. During Visit 1 (December 2012; \citealt{Kulow2014}), one exposure was taken before the optical transit, one during and two after. Visits 2 and 3 (June 2013 and 2014; \citealt{Ehrenreich2015}) have two exposures before the optical transit, one during and one after. The times of mid-exposures were calculated with the ephemeris in \citealt{Knutson2011}, and are reported in Table~\ref{obs_log}. 

\begin{table*}
\centering
%\begin{minipage}[b]{0.9\textwidth}
\begin{tabular}{lcccccc}
\hline
\hline
\noalign{\smallskip}  	
Time         &\multicolumn{2}{c}{Pre-transit}&Transit	&\multicolumn{2}{c}{Post-transit}	\\      
\noalign{\smallskip}
\hline
\noalign{\smallskip}
Visit 1 		&   -             & [-01:55 ; -01:30]      &   [-00:43 ; 00:05]  &   [00:52 ; 01:41]  &   [02:28 ; 03:17] \\ 
\noalign{\smallskip} 
Visit 2 		&[-03:23 ; -02:55]& [-02:01 ; -01:27]      &   [-00:26 ; 00:09]  &   [01:10 ; 01:45]  &   - \\ 
\noalign{\smallskip}
Visit 3 		&[-03:26 ; -02:58]& [-02:00 ; -01:25]      &   [-00:24 ; 00:10]  &   [01:11 ; 01:46]  &   - \\
\noalign{\smallskip}
\hline
\hline
\end{tabular}
\caption{Log of GJ\,436b transit observations. Time is given in hours and minutes, and counted from the center of the optical transit.}
\label{obs_log}
%\end{minipage}
\end{table*}

%%%%%%%%%%%%%%%%%%%%%%%%%%%%%%%%%%%%%%%%%%%%%%%%%%%%%%%%%%%%%%%%%%%%%%%%%%%%%%

\section{Reconstruction of the Lyman-$\alpha$ line}
\label{lalpha}

\subsection{Method}

The observed Lyman-$\alpha$ line is affected by absorption from the interstellar medium (ISM), which needs to be corrected for in order to estimate the intrinsic stellar Lyman-$\alpha$ line profile as seen from the planet. It is used to calculate the radiation pressure on the escaping exosphere (which is dependent on the stellar flux), and as a reference to calculate the theoretical transmission spectra of the exosphere as seen with HST/STIS (Sect.~\ref{sec:theo_spec}). The intrinsic line can be reconstructed directly using the STIS observations, and we applied the same method to all visits independently (\citealt{Wood2005}, \citealt{Ehrenreich2011}, \citealt{France2012}, \citealt{Bourrier2013}). We calculate a model profile of the stellar Lyman-$\alpha$ line, which is absorbed by the interstellar hydrogen and deuterium, convolved by STIS line spread function (LSF), and then compared with an observed reference spectrum. For each visit we used as reference the spectrum obtained during the first HST orbit, assuming it is unabsorbed by the exosphere of GJ\,436 b. We used the Bayesian Information Criterion (BIC; \citealt{deWit2012}) to prevent over-fitting in our search for the best model for the Lyman-$\alpha$ and LSF profiles. Trying out different models for the stellar line and the LSF, we found that in all visits the lowest BICs were obtained with a single Voigt profile for the stellar emission line, and a single Gaussian profile for the LSF. The free parameters of the fit are thus the heliocentric velocity of the star $\gamma_{*}$, the Lyman-$\alpha$ line profile parameters (maximum stellar flux at 1\,au f$_{peak}$(1\,au), Doppler width $\Delta$\,v$_{D}$ and damping parameter $a_{damp}$), the width of the LSF profile $\sigma_{LSF}$, and the ISM parameters for neutral hydrogen  (column density log$_{10}$\,$N$(H\,{\sc i}), Doppler broadening parameter $b$(H\,{\sc i}), and velocity relative to the star $\gamma$(H\,{\sc i})$_{/*}$). Although there may be multiple ISM components along the line of sight (LOS) toward GJ\,436, we assumed their absorption of the stellar Lyman-$\alpha$ profile can be interpreted as that of a single component with average properties. We thus modeled the ISM opacity as the combination of two Voigt profiles for the atomic hydrogen and deuterium, separated by about 0.33\,\AA. We used a D\,{\sc i}/H\,{\sc i} ratio of 1.5$\times$10$^{-5}$\ ({\it e.g.}, \citealt{Hebrard_Moos2003}, \citealt{Linsky2006}) and assumed there is no turbulent broadening with $b$(H\,{\sc i}) = $b$(D\,{\sc i})/$\sqrt{2}$ (\citealt{Wood2005}).    \\
A preliminary analysis of $\chi^2$ variations was used to locate the best model, and revealed no multiple local minima in the parameter space. We then refined the best-fit parameter values and calculated their uncertainties using a Metropolis-Hasting Markov chain Monte Carlo (MCMC) algorithm. To better sample the posterior distribution in the case of non-linear correlations between parameters, we applied an adaptive principal component analysis to the chains and jumped the parameters in an uncorrelated space (\citealt{diaz2014}). The step size was adjusted to ensure an acceptation rate of about 25\%, and the system was analyzed with chains of 5$\times$10$^{5}$ accepted steps. The comparison between the model spectra and the observed spectra was performed over the entire spectral line, excluding the core where ISM absorption is saturated and noise from the airglow correction is high (the spectral range of the fits are given in Table~\ref{results_la}). The medians of the posterior probability distributions were chosen as the final best-fit values for the model parameters, and their 1$\sigma$ uncertainties were obtained by finding the intervals on both sides of the median that contain 34.15\% of the accepted steps. 

\subsection{Results}
\label{la_results}

\begin{figure}     %haut   gauche    bas droite
\includegraphics[angle=-90,trim=4.cm 4cm 4.9cm 1cm,clip=true,width=\columnwidth]{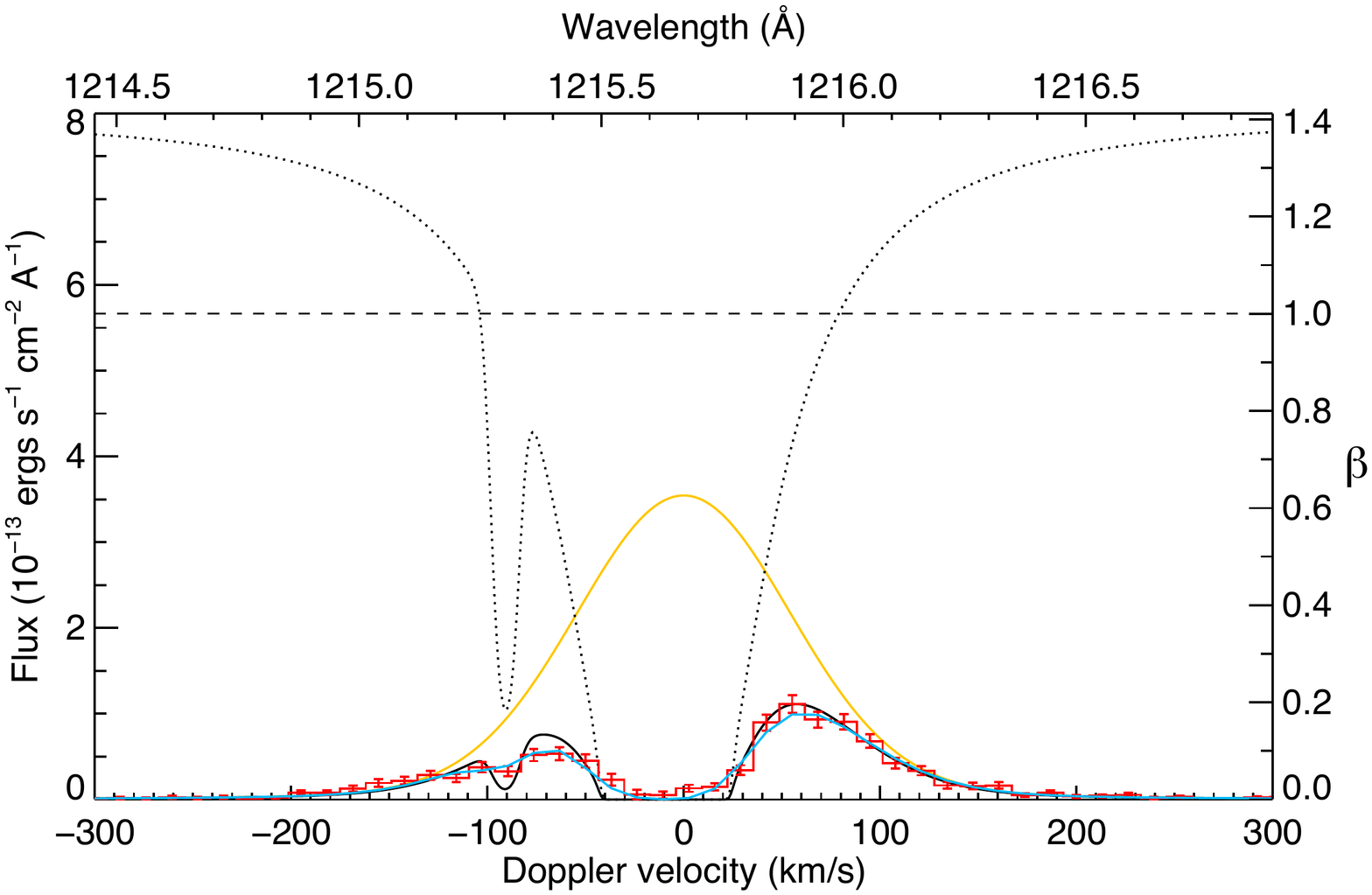}
\includegraphics[angle=-90,trim=4.cm 4cm 3cm 1cm,clip=true,width=\columnwidth]{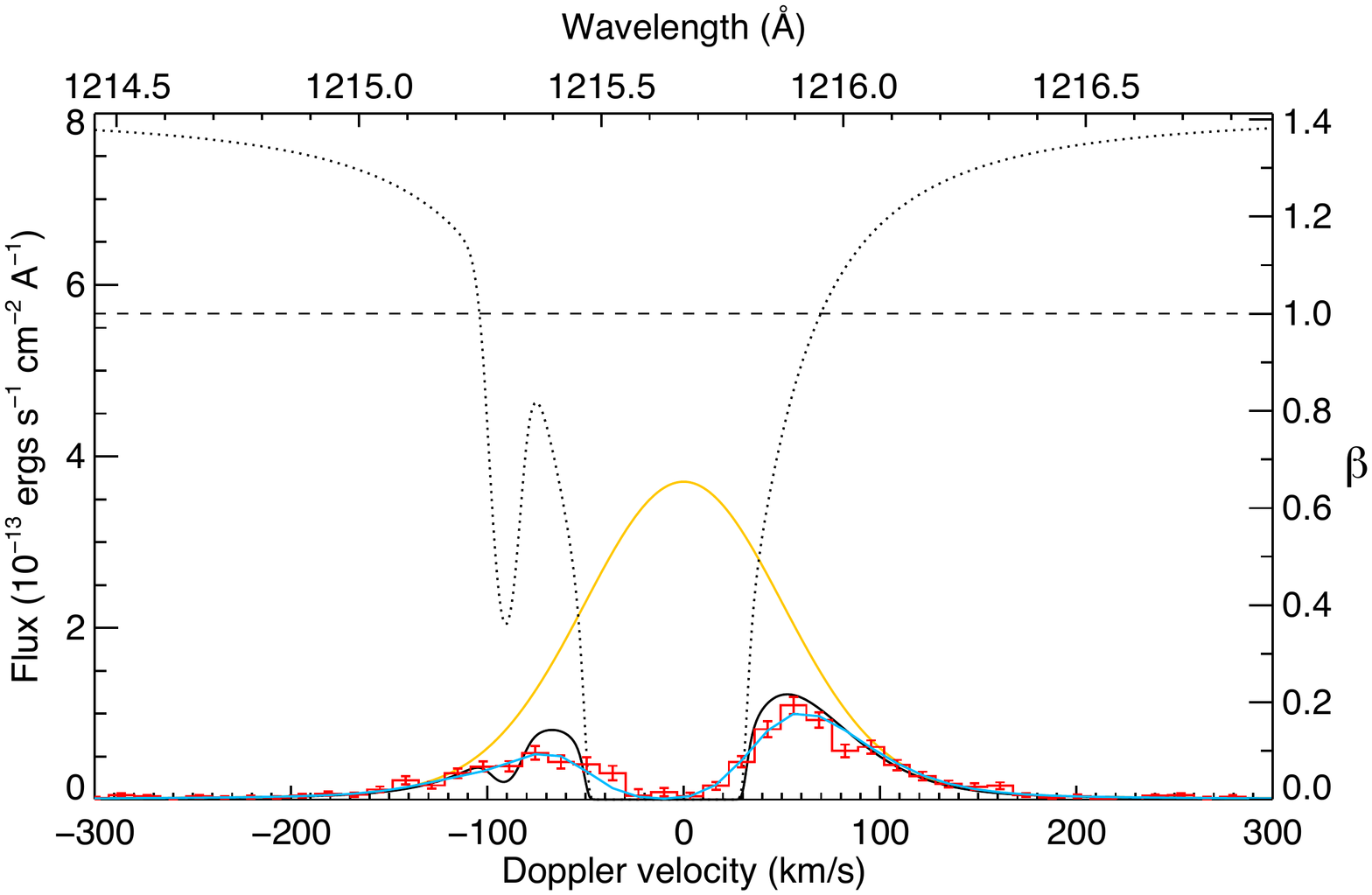}
\caption[]{Lyman-$\alpha$ line profiles of GJ\,436 for Visits 2 (top) and 3 (bottom). The yellow line shows the theoretical intrinsic stellar emission line profile as seen by hydrogen atoms escaping the planetary atmosphere, scaled to the Earth distance. It also corresponds to the profile of the ratio $\beta$ between radiation pressure and stellar gravity (values reported on the right axis). The solid black line shows the Lyman-$\alpha$ line profile after absorption by the interstellar hydrogen (1215.6\AA) and deuterium (1215.25\AA), whose cumulated profile is plotted as a dotted black line (ISM absorption in the range 0-1 has been scaled to the entire vertical axis range). The line profile convolved with the LSF (blue line) is compared to the observations shown as a red histogram. Above the horizontal dashed line ($\beta$=1), radiation pressure would overcome stellar gravity and accelerate particles away from the star. However with $\beta$ values below unity, radiation pressure from GJ\,436 can only brake the gravitational deviation of particles toward the star.}
\label{fits_lalpha_GJ436}
\end{figure}

The best-fit reconstructed Lyman-$\alpha$ stellar profiles are displayed in Fig.~\ref{fits_lalpha_GJ436}, with the corresponding model parameters given in Table~\ref{results_la}. The reconstructed Lyman-$\alpha$ lines for Visits 2 and 3 are remarkably similar, with all fit parameters consistent within their 1$\sigma$ uncertainties. On the other hand we found spurious values for the best-fit parameters in Visit 1, with the ISM absorption profile anomalously deep and strongly blue-shifted. This is consistent with a lower flux in the blue wing of the Lyman-$\alpha$ line caused by absorption from hydrogen escaping the planet, and confirms that the spectrum obtained during the first exposure in Visit 1 is already absorbed by the giant exosphere of GJ 436 b (\citealt{Ehrenreich2015}). Without a good knowledge of both wings of the line, it is not possible to reconstruct the intrinsic stellar line for Visit 1. However the atmospheric absorption occurs in a localized spectral range in the blue wing of the line, and \citealt{Ehrenreich2015} showed that the Lyman-$\alpha$ flux measured at larger velocity blue shifts, and over the whole red wing of the line, is very stable between the three visits, which is in agreement with our line reconstructions in Visits 2 and 3. Hereafter we thus use the Lyman-$\alpha$ line profile obtained for Visit 2 as a proxy for Visit 1. \\
The best-fit parameters we obtained for Visits 2 and 3 correspond to heliocentric radial velocity for the ISM hydrogen of -1.7$\pm$2.1\,km\,s$^{-1}$ and -1.0$\pm$2.7\,km\,s$^{-1}$, respectively. Using the LISM Kinematic Calculator\footnote{\mbox{\url{http://sredfield.web.wesleyan.edu/}}} (\citealt{Redfield_Linsky2008}), we identified three ISM clouds near (less than 20$^\circ$) the line-of-sight toward GJ\,436 with radial velocities of 0.37$\pm$1.38\,km\,s$^{-1}$ (LIC cloud), 2.33$\pm$0.76\,km\,s$^{-1}$ (Leo cloud), and 5.81$\pm$0.74\,km\,s$^{-1}$ (NGP cloud). The ISM absorption derived from our reconstruction of the Lyman-$\alpha$ line thus corresponds more likely to the LIC cloud ($<$1$\sigma$), and may be consistent with the Leo cloud ($<$2$\sigma$).
							
\begin{table*}
%\centering
\caption{Parameters derived from the reconstruction of the Lyman-$\alpha$ line profile in Visits 2 and 3. The line could not be reconstructed in Visit 1 due to the lack of reference spectrum.}
\begin{tabular}{llccl}
\hline
\noalign{\smallskip}  
 &   \textbf{Parameter}         & \textbf{Visit 2}         & \textbf{Visit 3} 		   & \textbf{Unit} \\   			
\noalign{\smallskip}
\hline
\hline
\noalign{\smallskip}
\textit{Fit range}&  Blue wing  		    	   		  & [1214.64 - 1215.53] 		    &[1214.68 - 1215.53]   & \AA\     \\ 
     		     &  Red wing  		        		  & [1215.75 - 1216.70] 			&[1215.75 - 1216.54]$^\dagger$   & \AA\     \\ 
\textit{ISM (H\,{\sc i})} &  log$_{10}$\,$N$(H\,{\sc i}) &  18.01$\pm$0.07               &17.90$\stackrel{+0.15}{_{-0.19}}$& cm$^{-2}$\\
                 &  $b$(H\,{\sc i})    &  9.9$\stackrel{+1.3}{_{-1.7}}$&12.3$\stackrel{+1.5}{_{-2.1}}$	 &km\,s$^{-1}$\\
        		&  $\gamma$(H\,{\sc i})$_{/*}$  & -9.1$\stackrel{+1.4}{_{-1.3}}$&-9.2$\stackrel{+1.7}{_{-1.9}}$ 	 &km\,s$^{-1}$\\
\textit{LSF}   	&  $\sigma_{LSF}$     &  0.97$\pm$0.14            	    &1.25$\stackrel{+0.18}{_{-0.17}}$  & HST pixels\\
\textit{Stellar line}&   f$_{peak}$(1\,au)   & 1.55$\stackrel{+0.34}{_{-0.24}}$&1.62$\stackrel{+0.56}{_{-0.38}}$  & erg cm$^{-2}$ s$^{-1}$ \AA$^{-1}$\\    
                 & $\Delta$\,v$_{D}$           & 74$\pm$3  			     &69$\pm$5         &  km\,s$^{-1}$	\\		                            
                 &  $a_{damp}$  			   &  0.09$\pm$0.02 			      & 0.10$\stackrel{+0.03}{_{-0.02}}$  	&  	\\
                 &  $\gamma_{*}$    &  7.4$\pm$1.6                    & 8.2$\stackrel{+2.1}{_{-1.7}}$  	& km\,s$^{-1}$  \\
				&   F$_{Ly\alpha}$(Earth)  & 2.05$\times$10$^{-13}$ & 2.01$\times$10$^{-13}$ & erg cm$^{-2}$ s$^{-1}$\\                 
				    &  F$_{Ly\alpha}$(1\,au)   & 0.90 				   & 0.88  & erg cm$^{-2}$ s$^{-1}$\\                 
\hline									
\hline
\multicolumn{5}{l}{Note: F$_{Ly\alpha}$ is the flux integrated in the entire Lyman-$\alpha$ line and calculated at Earth distance and at 1\,au from the star.} \\
\multicolumn{5}{l}{$\dagger$: An anomalous low-flux pixel at about 1206.0\AA\ was removed from the fit (see Fig.~\ref{fits_lalpha_GJ436}).} \\
\end{tabular}
\label{results_la}
\end{table*}

\section{\textit{EVE}: the EVaporating Exoplanets code}
\label{model}

EVE is a 3D numerical code that we developed to calculate the structure of an exoplanet upper atmosphere and the transmission spectrum of the species it contains. The direct comparison of these spectra with observational data allows the inferrance of several physical properties of both the planetary atmosphere and the star, and the code has been used in this way to study the neutral hydrogen exospheres of the hot Jupiters HD\,209458b and HD\,189733b (\citealt{Bourrier_lecav2013}), the warm Jupiter 55 Cnc b, and the super-Earth 55 Cnc e (\citealt{Ehrenreich2012}), as well as the neutral and ionized magnesium populations in the thermosphere/exosphere of HD\,209458b (\citealt{Bourrier2014}). The EVE code was used in \citet{Ehrenreich2015} to perform a preliminary fit of the three combined observations of GJ\,436b neutral hydrogen exosphere. In the present paper, we further investigate the effects of stellar radiation pressure, stellar photo-ionization, and the planetary escape properties on the structure of the exosphere and its transmission spectrum. The following subsections offer a synthetic description of the model used in this frame.  

\subsection{Bottom layers of the atmosphere}

	The upper atmosphere is divided in two different regimes, joined at a transition altitude set at $R_{\mathrm{trans}}$=4\,$R_{\mathrm{planet}}$, approximately the mean altitude of the Roche lobe. We assume that the atmosphere within the Roche lobe, which is by definition under the gravitational influence of the planet, keeps a global cohesion and can be described analytically with a radial density profile corresponding to a hydrostatic equilibrium. In the numerical simulation, the density profile of neutral hydrogen in this part of the atmosphere is reassessed every time step by adjusting its value at $R_{\mathrm{trans}}$ to match the hydrogen density obtained in the upper atmospheric layers (see Sect.~\ref{sec:upp}). Because of the eccentricity of the planetary orbit (e=$0.16$), the density profile thus varies over time. The gas filling the Roche Lobe mainly contributes to the absorption in the core of the Lyman-$\alpha$ line, which cannot be observed because of ISM absorption and geocoronal emission. Therefore, this part of the atmosphere has little influence on the observed spectra, except through the Lorentzian wings of the absorption profile (the so-called damping wings, e.g., \citealt{BJ2008}). This is however excluded because no significant absorption signal was observed in the red wing of the Lyman-$\alpha$ line during the optical transit of GJ\,436 b (\citealt{Ehrenreich2015}).\\
		
\subsection{Upper layers of the atmosphere}
\label{sec:upp}

We use Monte-Carlo particle simulations to compute the dynamics of the escaping gas above $R_{\mathrm{trans}}$. Neutral hydrogen atoms are represented by meta-particles, and the total number of metaparticles launched every time step $dt$, set to 5\,mn, depends on the escape rate of neutral hydrogen along the eccentric orbit. To take the eccentricity into account, we assumed that GJ\,436b is in the energy-limited regime of evaporation, and the escape rate is proportional to the stellar energy input into the atmosphere. We thus varied the escape rate as the inverse square of the distance to the star, setting the reference $\dot{M}_{\mathrm{H^{0}}}$ at the semi-major axis. The particles are released from the entire atmosphere and their initial velocity distribution, relatively to the planet, is the combination of the upward bulk velocity of the planetary wind $v_{\mathrm{pl-wind}}$ and an additional thermal speed component from a Maxwell-Boltzman velocity distribution (see \citealt{Bourrier_lecav2013}).\\
The dynamics of the particles is calculated in the stellar reference frame and results from the stellar and planetary gravities, the stellar radiation pressure, and the inertial force linked to the non-Galilean reference frame. Radiation pressure on an atom is the combination of a radial\footnote{Hereafter, we use the term \textit{projected velocity} to refer to the projection of a particle velocity on the star-Earth line of sight, and \textit{radial velocity} to refer to its projection on the star-particle axis. Radiation pressure varies with the radial velocity of a particle, while we measure spectra as a function of projected velocity.} impulsion in the direction opposite to the star, caused by the absortion of a photon, followed by an isotropic impulsion after the photon reemission. We process the cumulated radial impulses from all photons absorbed by a hydrogen atom during $dt$ as a force proportional to stellar gravity (\citealt{Lagrange1998}), and apply an equivalent number of velocity changes with constant speed 3.26\,m\,s$^{-1}$ (this value comes from the conservation of momentum) and random direction to reflect the isotropic impulsions. A crucial aspect of radiation pressure is its dependence on a hydrogen atom radial velocity, as the number of photons it absorbs and reemits is proportional to the flux in the Lyman-$\alpha$ line Doppler-shifted at this velocity. The effect of radiation pressure thus strongly depends on the shape of the line profile, and the single peak profile obtained for GJ\,436b (Fig.~\ref{fits_lalpha_GJ436}) implies that the strength of radiation pressure monotonously decreases from the core to the wings of the line, ie with increasing absolute radial velocities.\\
The hydrogen atoms lifetime depends on the stellar photo-ionization rate $\Gamma_{\mathrm{ion}}$ which varies as the inverse square of the distance to the star. Once ionized, hydrogen atoms are removed from the simulation.\\
We do not take collisions into account in the upper atmospheric layers, as neutral hydrogen densities reach about 4$\times$10$^{5}$\,m$^{-3}$ at $R_{\mathrm{trans}}$ for our best-fit simulations. With our simulation settings this value corresponds to a Knudsen number (the ratio between the mean free path of the gas and the atmospheric scale height) of about 40, which is above the level of the exobase defined by a Knudsen of one, even if the neutral fraction of hydrogen in the upper atmosphere of GJ\,436 b is about 0.1 (\citealt{Kulow2014}, \citealt{Koskinen2013a}).

\subsection{Theoretical spectra}
\label{sec:theo_spec} 

Self-shielding for radiation pressure and photo-ionization is taken into account by calculating their respective opacities in the bottom and upper atmospheric layers. Similarly, EVE calculates the opacity along the star-Earth line of sight to obtain at each time step the theoretical absorption spectra in the Lyman-$\alpha$ line (for details regarding the calculation of these opacities, see \citealt{Bourrier_lecav2013}). During the optical transit, the stellar flux is absorbed at every wavelength by the planetary disk. In contrast, the neutral hydrogen atmosphere may contribute to the absorption on longer time scales and at specific wavelengths that depends on the dynamics of the gas and the effects of thermal and natural broadening. At a given time step in the simulation, the theoretical intrinsic stellar line profile is multiplied by the absorption profile of the planet and its atmosphere, calculated with a resolution $\Delta v=$10\,km\,s$^{-1}$ (more than twice the resolution of the STIS spectra). The resulting spectrum is then attenuated by the ISM absorption and convolved with the LSF (using the fits obtained in Sect.~\ref{la_results}), and interpolated to the observed velocities. To compare the simulation results with the STIS observations, we average the theoretical spectra calculated during the time window of each observation. \\

\section{Structure of the exosphere}
\label{struct}

While mechanisms such as stellar wind or bow-shocks may play a role in shaping the exosphere of an exoplanet, their description requires strong assumptions and the introduction of several free parameters in models of atmospheric escape. By contrast radiation pressure is known to be present with a well defined behavior, and we showed in Sect.~\ref{lalpha} that it is very well constrained from direct measurements of the stellar Lyman-$\alpha$ line. In this paper we thus focus on the effects of radiation pressure alone on the structure and the spectral signature of the exosphere. 

\subsection{Radiation pressure: shaping the cloud}
\label{struct_rad_brak}

\begin{figure}     %gauche   bas    droite haut
\centering
\includegraphics[trim=2.4cm 3.3cm 6cm 12.5cm,clip=true,width=\columnwidth]{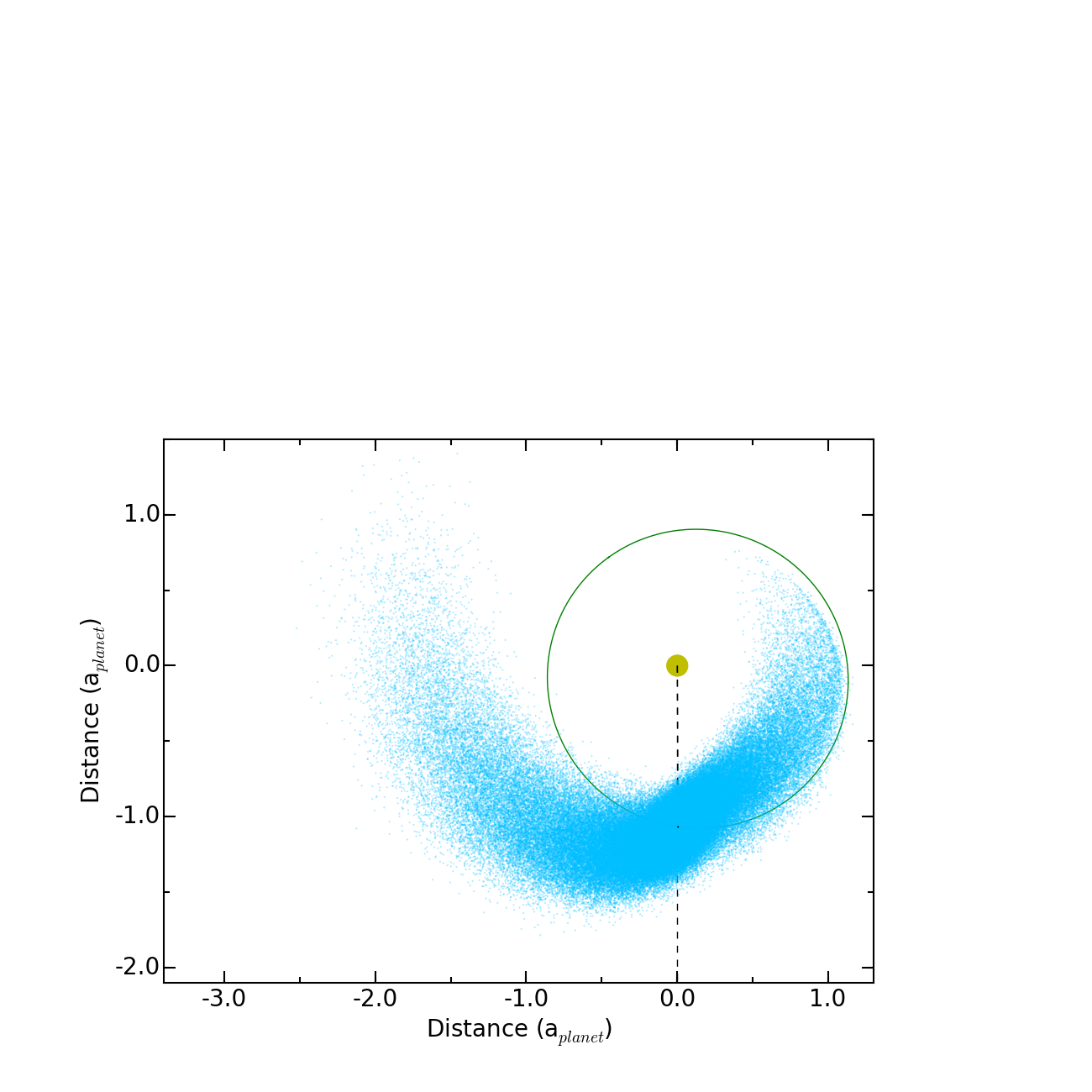}
\includegraphics[trim=2.4cm 3.3cm 6cm 13cm,clip=true,width=\columnwidth]{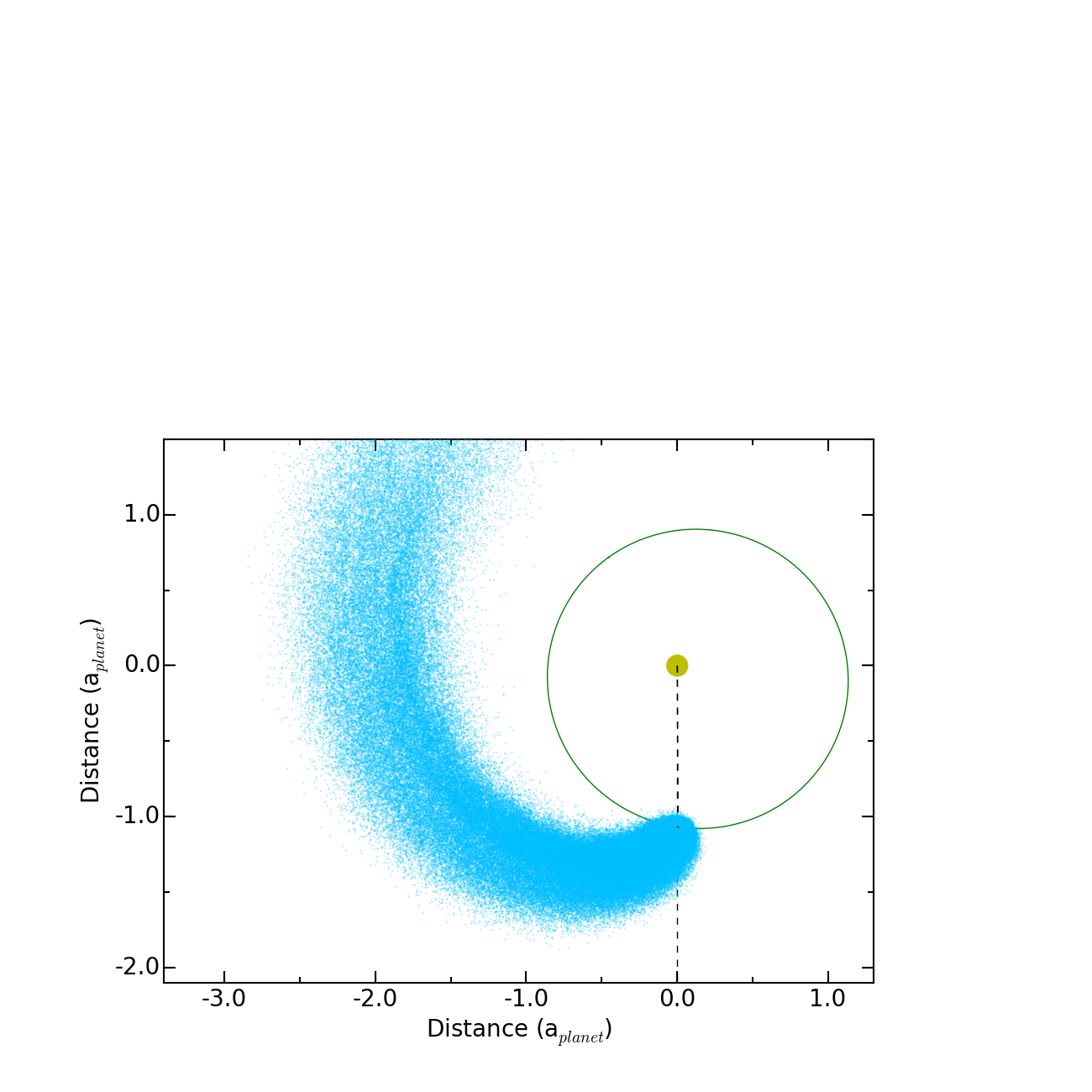}
\includegraphics[trim=2.4cm 1cm 6cm 13cm,clip=true,width=\columnwidth]{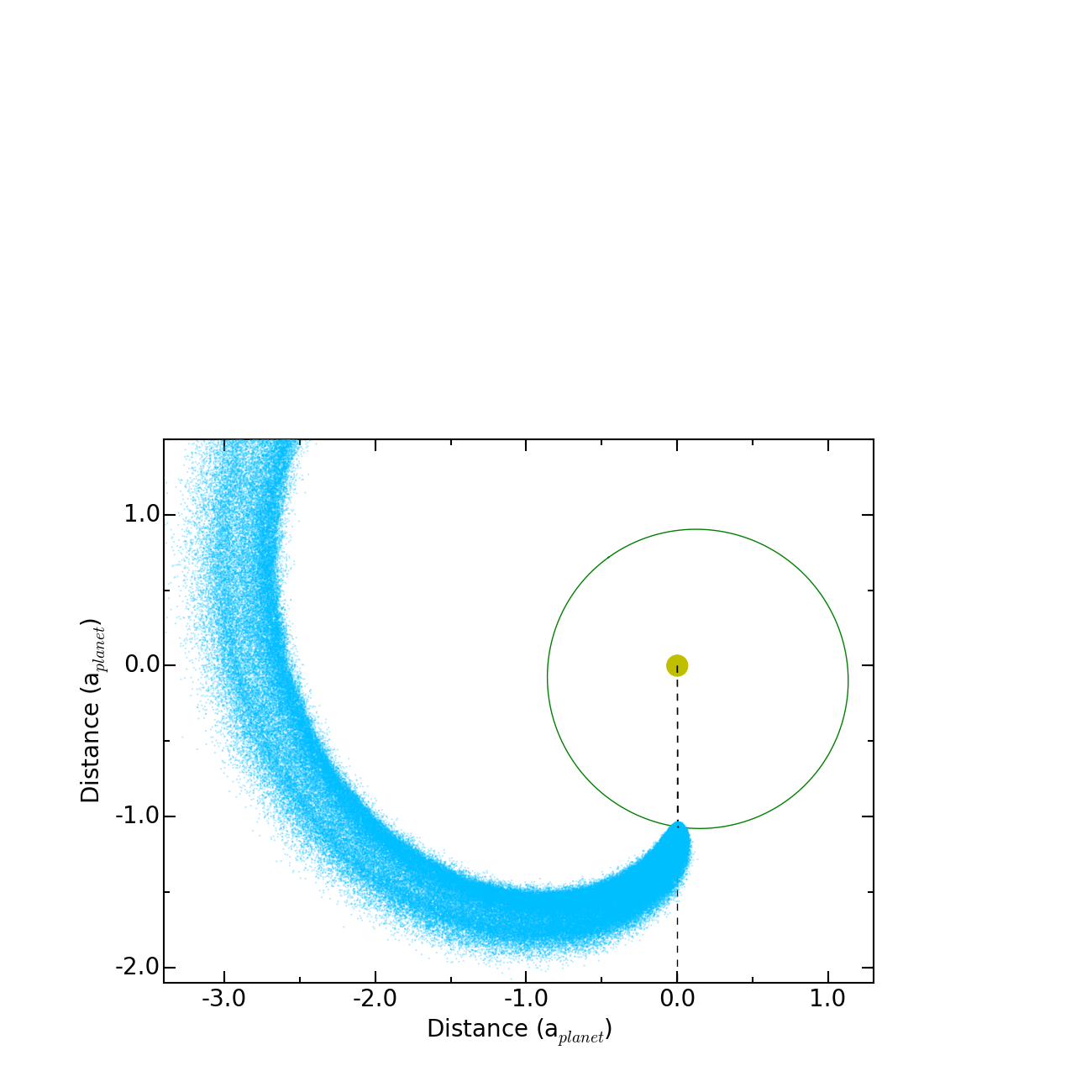}
\caption[]{Views of GJ\,436b exosphere (blue dots) from the perpendicular to the orbital plane. The planet is the small disk at the intersection of the star/Earth LOS (dashed black line) and the planetary orbit (green line). All system properties are similar in the three panels, except for radiation pressure. \textbf{Upper panel:} Without radiation pressure, stellar gravity shears the extended cloud of hydrogen, and gas at shorter orbital distance than the planet falls toward the star. \textbf{Middle panel:} With even a moderate radiation pressure ($\sim$60\% of stellar gravity at maximum), all the escaping gas decelerates and moves to larger orbits, forming a longer comet-like tail trailing behind the planet. This is the real radiation pressure that corresponds to the observed Lyman-$\alpha$ flux. \textbf{Lower panel:} With a high radiation pressure ($\sim$400\% of stellar gravity at maximum), the escaping gas is swiftly blown away. The planet is surrounded by a much smaller coma but trailed by a narrower, more radial, cometary-tail.}
\label{cloud_rad_brak}
\end{figure}

If the atmosphere of an exoplanet is significantly expanded, various regions of the exosphere move on very different orbits and revolve around the star in differential rotation. In the absence of any radiation pressure from the star, gas escaping beyond the Roche Lobe would follow ballistic trajectories subjected to stellar gravity alone. The exosphere would suffer shear, the gas within the orbit of the planet accelerating and falling toward the star while the gas beyond would decelerate and move toward larger distances (top panel in Fig.~\ref{cloud_rad_brak}). However, this is an unrealistic case for neutral hydrogen escape, as radiation pressure from the Lyman-$\alpha$ line will always compensate for some fraction of the stellar gravity. In the case of close-in planets around K-type stars and earlier, such as HD\,189733b and HD\,209458b, the Lyman-$\alpha$ line is actually strong enough for radiation pressure to overcome up to 3-5 times the gravity, thus swiftly accelerating the H$^{0}$ atoms away from the star at radial velocities exceeding 100\,km\,s$^{-1}$ (bottom panel in Fig.~\ref{cloud_rad_brak}). The exosphere is compressed on its dayside,  while gas escaping at the limbs of the atmosphere is blown away and shaped into a narrow cometary tail, bent toward the radial direction of the stellar photons (e.g., \citealt{Schneider1998}, \citealt{VM2003}, \citealt{Bourrier_lecav2013}). Orbiting a M dwarf at moderate orbital distance, the warm Neptune GJ\,436 b comparatively receives a lower flux from the stellar Lyman-$\alpha$ line. With radiation pressure compensating for up to $\sim$60\% of the gravitational attraction at low radial velocities (Fig.~\ref{fits_lalpha_GJ436}), the escaping hydrogen atoms cannot be accelerated away from the planet but are nonetheless subjected to a strong \textit{radiative braking}. Indeed, even a small reduction of stellar gravity is enough for all the exospheric gas to decelerate and move to larger orbits, preventing the formation of a stream of matter infalling toward the star. In contrast to highly irradiated planets, the gas can diffuse within a larger volume around the planet and forms a wider cometary tail closer to the planet orbit (middle panel in Fig.~\ref{cloud_rad_brak}). \\

%version referee

%\begin{figure}     %gauche   bas    droite haut
%\centering
%\includegraphics[trim=2.4cm 3.3cm 6cm 12.5cm,clip=true,width=0.5\columnwidth]{Cloud_GJ436b_SANS_PRAD}
%\includegraphics[trim=2.4cm 3.3cm 6cm 13cm,clip=true,width=0.5\columnwidth]{Cloud_GJ436b_AVEC_PRAD}
%\includegraphics[trim=2.4cm 1cm 6cm 13cm,clip=true,width=0.5\columnwidth]{Cloud_GJ436b_AVEC_PRAD4}
%\caption[]{Views of GJ\,436b exosphere (blue dots) from the perpendicular to the orbital plane. The planet is the small disk at the intersection of the star/Earth LOS (dashed black line) and the planetary orbit (green line). All system properties are similar in the three panels, except for radiation pressure. \textbf{Upper panel:} Without radiation pressure, stellar gravity shears the extended cloud of hydrogen, and gas at shorter orbital distance than the planet falls toward the star. \textbf{Middle panel:} With even a moderate radiation pressure ($\sim$60\% of stellar gravity at maximum), all the escaping gas decelerates and moves to larger orbits, forming a longer comet-like tail trailing behind the planet. This is the real radiation pressure that corresponds to the observed Lyman-$\alpha$ flux. \textbf{Lower panel:} With a high radiation pressure ($\sim$400\% of stellar gravity at maximum), the escaping gas is swiftly blown away. The planet is surrounded by a much smaller coma but trailed by a narrower, more radial, cometary-tail.}
%\label{cloud_rad_brak}
%\end{figure}

%%%%%%%%%%%%%%%%%%%%%%%%%%%%%%%%%%%%%%%%%%%%%%%%%%%%%%%%%%%%%%%%%%%%%%%%%%%%%%%%

\subsection{Effect of radiation pressure on the velocity field}
\label{dyn_low_prad}

\begin{figure*}
\centering
\begin{minipage}[b]{0.9\textwidth}
\includegraphics[trim=2cm 1cm 2cm 8cm,clip=true,width=\columnwidth]{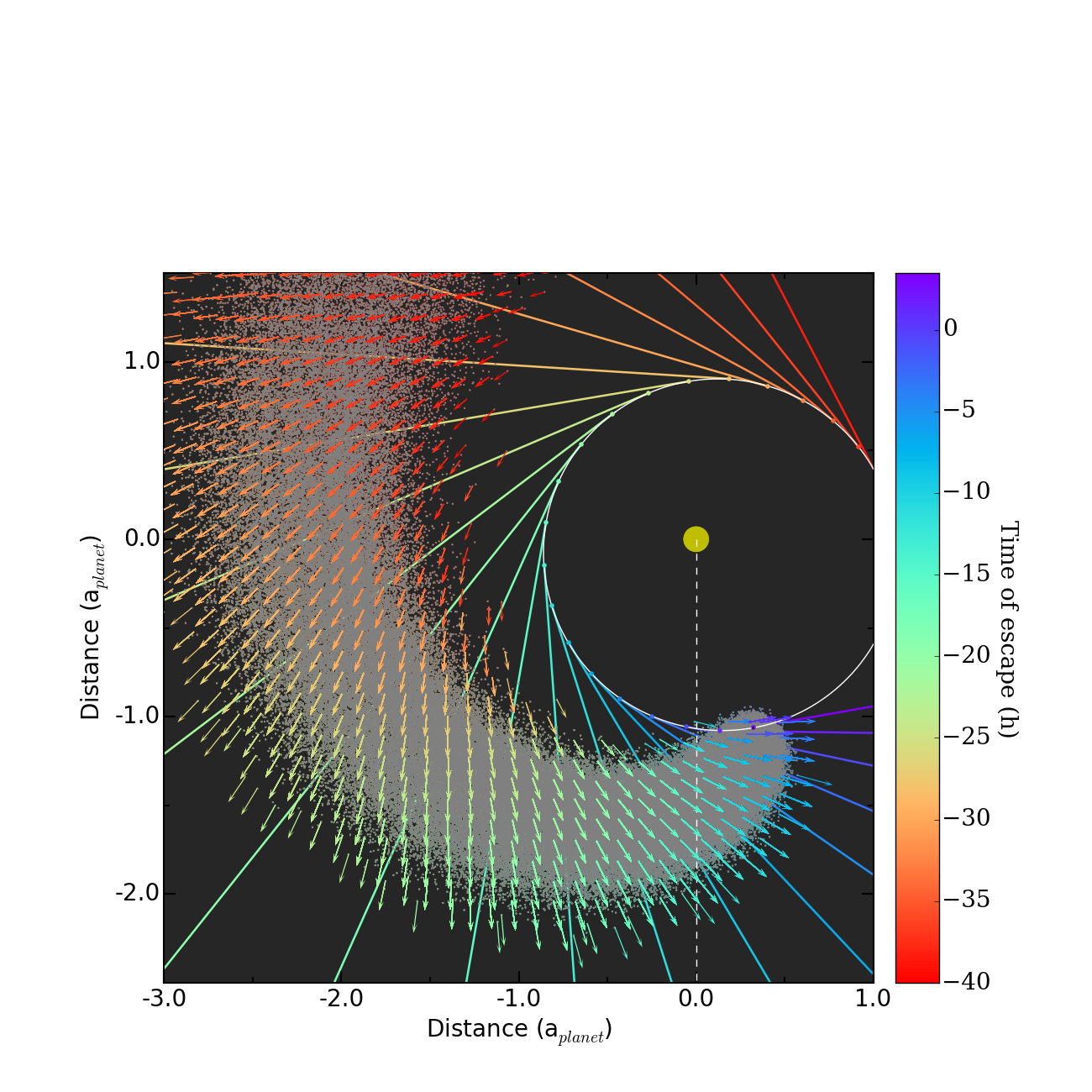}
\end{minipage}
\caption[]{Views of GJ\,436b exosphere (gray dots) from the perpendicular to the orbital plane, 3.5\,h after the optical transit. The white dashed line shows the LOS toward Earth. Arrows display the velocity field of the hydrogen atoms with respect to the star, colored as a function of the time they escaped the atmosphere (counted from the center of the optical transit). In the reduced gravity field from the star, particles move beyond the orbit of the planet. However, radiation pressure does not overcome stellar gravity, and particles are still deviated from the tangent to the orbit at the time of their escape (solid colored lines). Because radiation pressure is velocity-dependent, the particles do not stay on Keplerian orbits. To better illustrate the effect of radiative braking, this simulation was performed with a low ionization rate artificially lengthening the exospheric tail.}
\label{velfield_rad_brak}
\end{figure*}

Radiation pressure has also a major effect on the velocities of the escaping hydrogen atoms that correspond to the wavelength range of the observed absorption profile. In the stellar reference frame, the initial velocity of an escaping particle is the combination of the planet orbital velocity and the velocity of the planetary wind at the Roche Lobe. For close-in exoplanets ($a<$0.1\,au), orbital velocity is generally higher than 100\,km\,s$^{-1}$, significantly higher than planetary wind velocity in the order of 1-25\,km\,s$^{-1}$ with respect to the planet (\citealt{Yelle2004}; \citealt{Koskinen2013a}; \citealt{Bourrier2014}). To a first approximation the average dynamics of the gas in the exosphere is thus set by the planet orbital velocity $v_{pl}$. With even a low radiation pressure, hydrogen atoms decelerate with respect to the planet. Because radiation pressure does not fully overcome the star gravity, hydrogen atoms nevertheless remain attracted toward the star and will not move beyond the plane tangent to the planet orbit at the time of their escape, as illustrated in Fig.~\ref{velfield_rad_brak}. As a result of both the orbital velocity of the planet and the effect of radiation pressure, the projected velocity of the gas thus ranges approximately between the projection of $v_{pl}$ (for atoms close to the planet) and -135\,km\,s$^{-1}$, which corresponds to the highest planet orbital velocity. Therefore, this dynamics is consistent with the detection of blue-shifted absorption up to about -120\,km\,s$^{-1}$ during the transit of GJ\,436 b exosphere (\citealt{Ehrenreich2015}).\\

\begin{figure*}
%\centering
\begin{minipage}[b]{\textwidth}	  % gauche , bas, droite , haut
\hspace{0.017\textwidth}
\includegraphics[trim=0.4cm 0.1cm 0.8cm 0.cm,clip=true,width=0.317\textwidth]{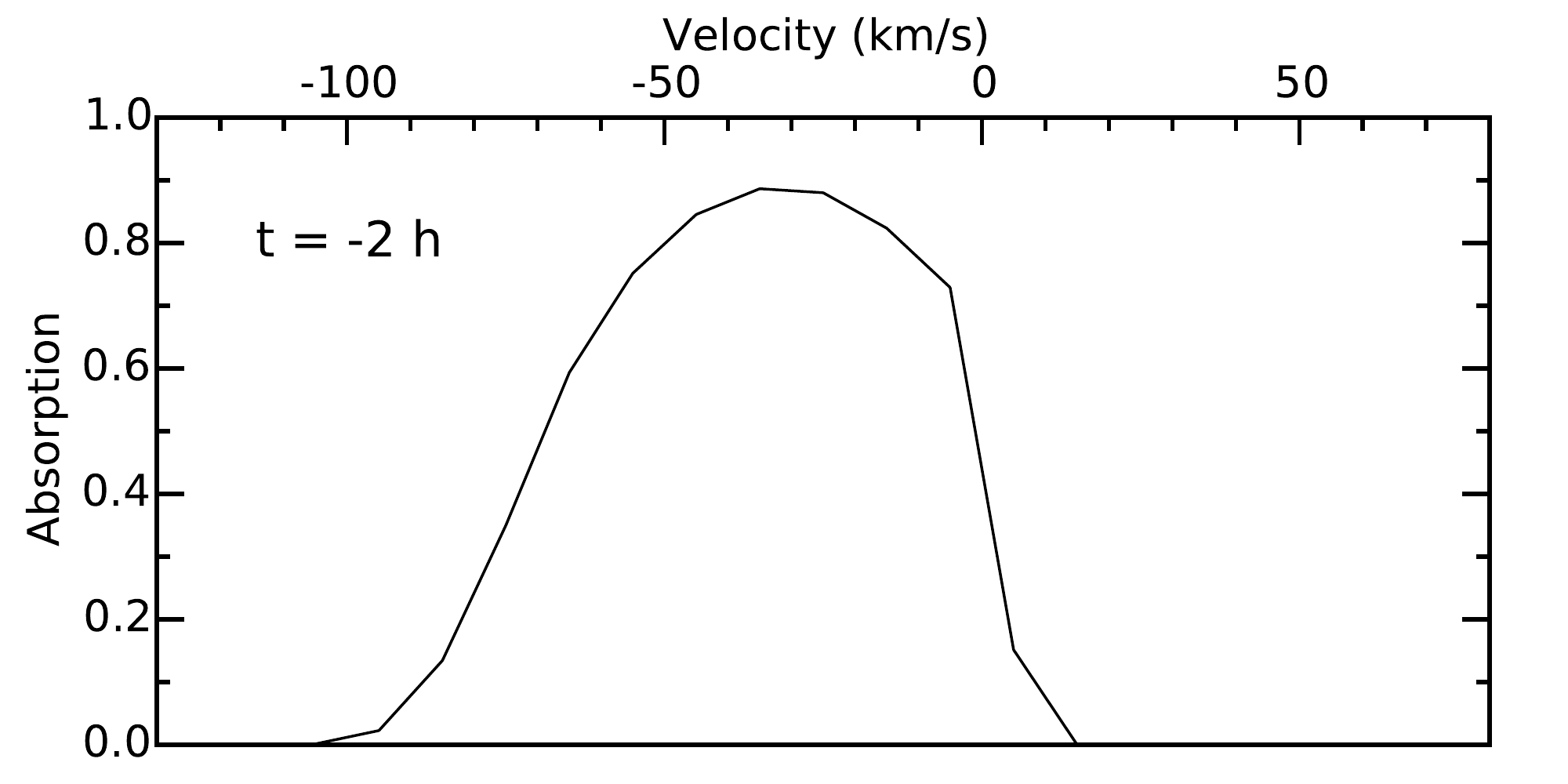}
\includegraphics[trim=2cm 0.1cm 0.8cm 0.cm,clip=true,width=0.291\textwidth]{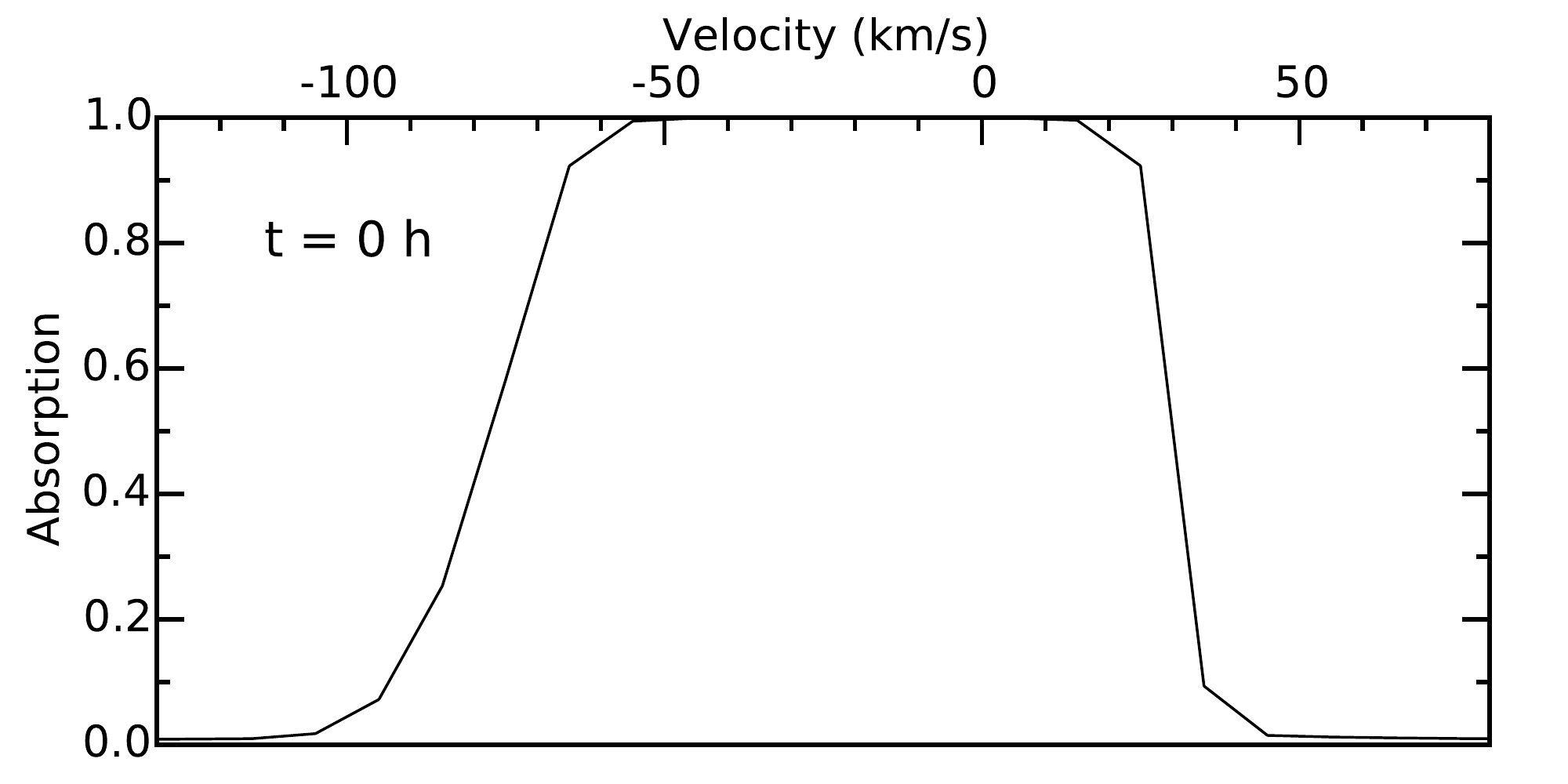}
\includegraphics[trim=2cm 0.1cm 0.8cm 0.cm,clip=true,width=0.291\textwidth]{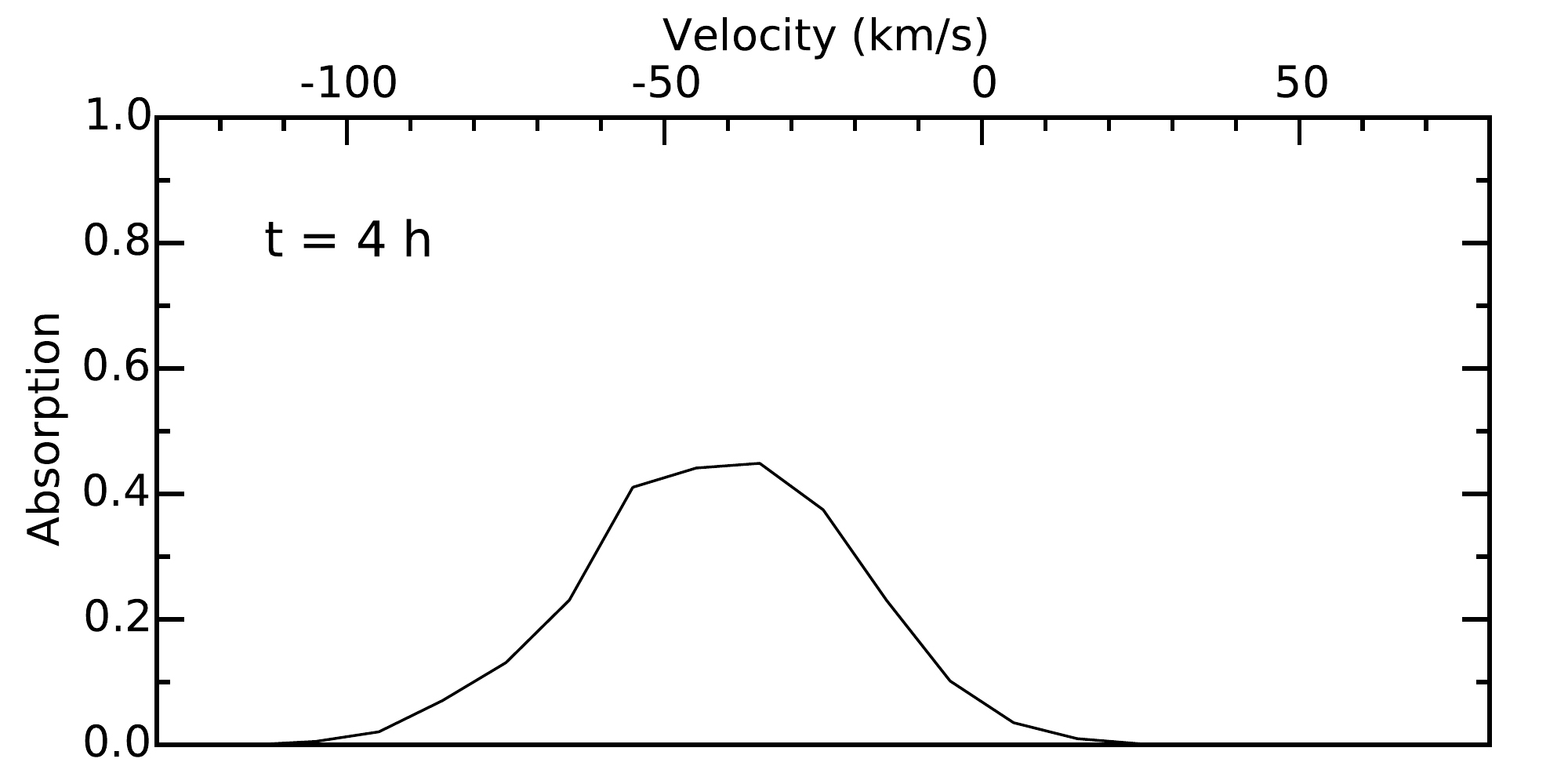}\\
\includegraphics[trim=1.2cm 1.5cm 6.4cm 7.85cm,clip=true,width=0.34\textwidth]{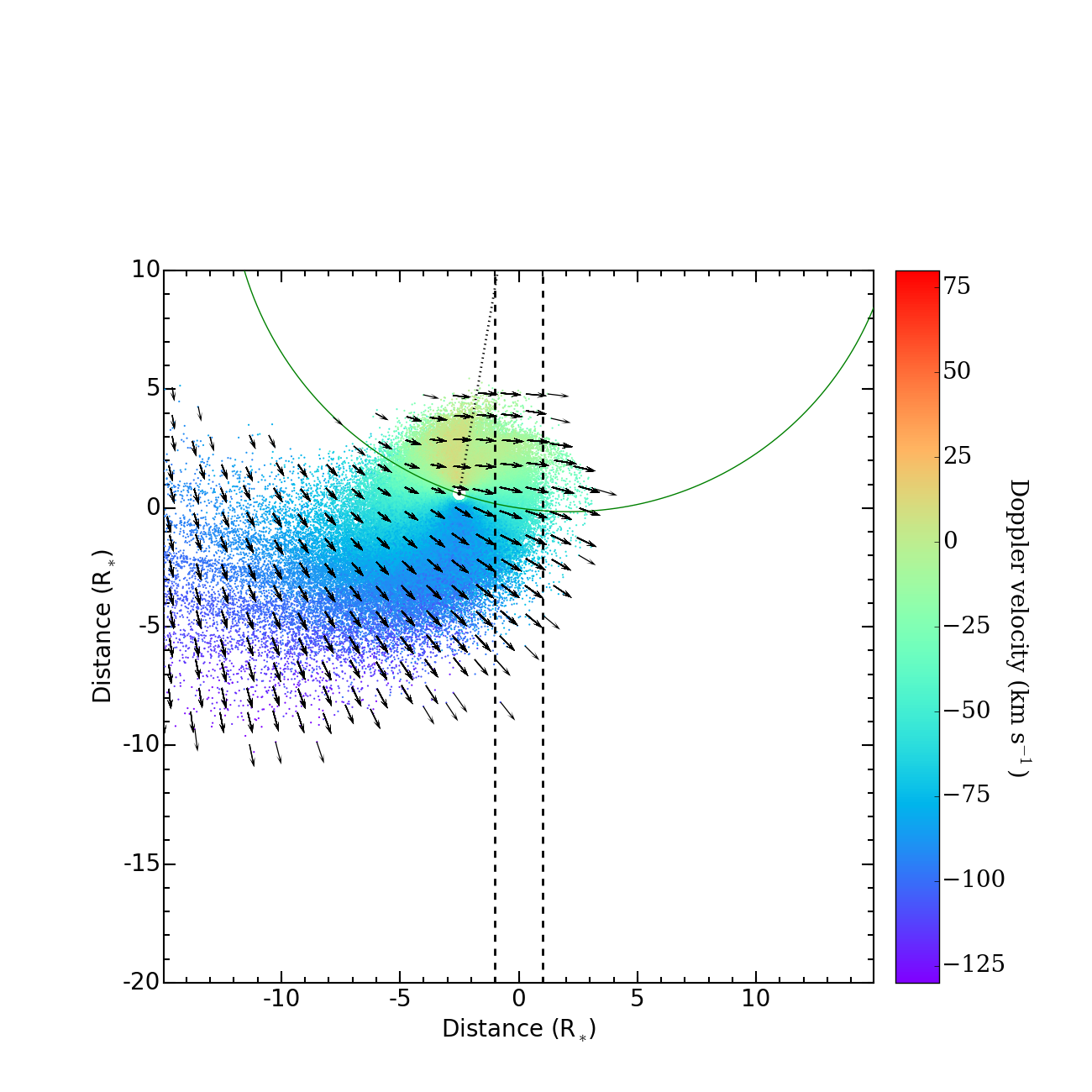}
\includegraphics[trim=4.85cm 1.5cm 6.4cm 7.85cm,clip=true,width=0.291\textwidth]{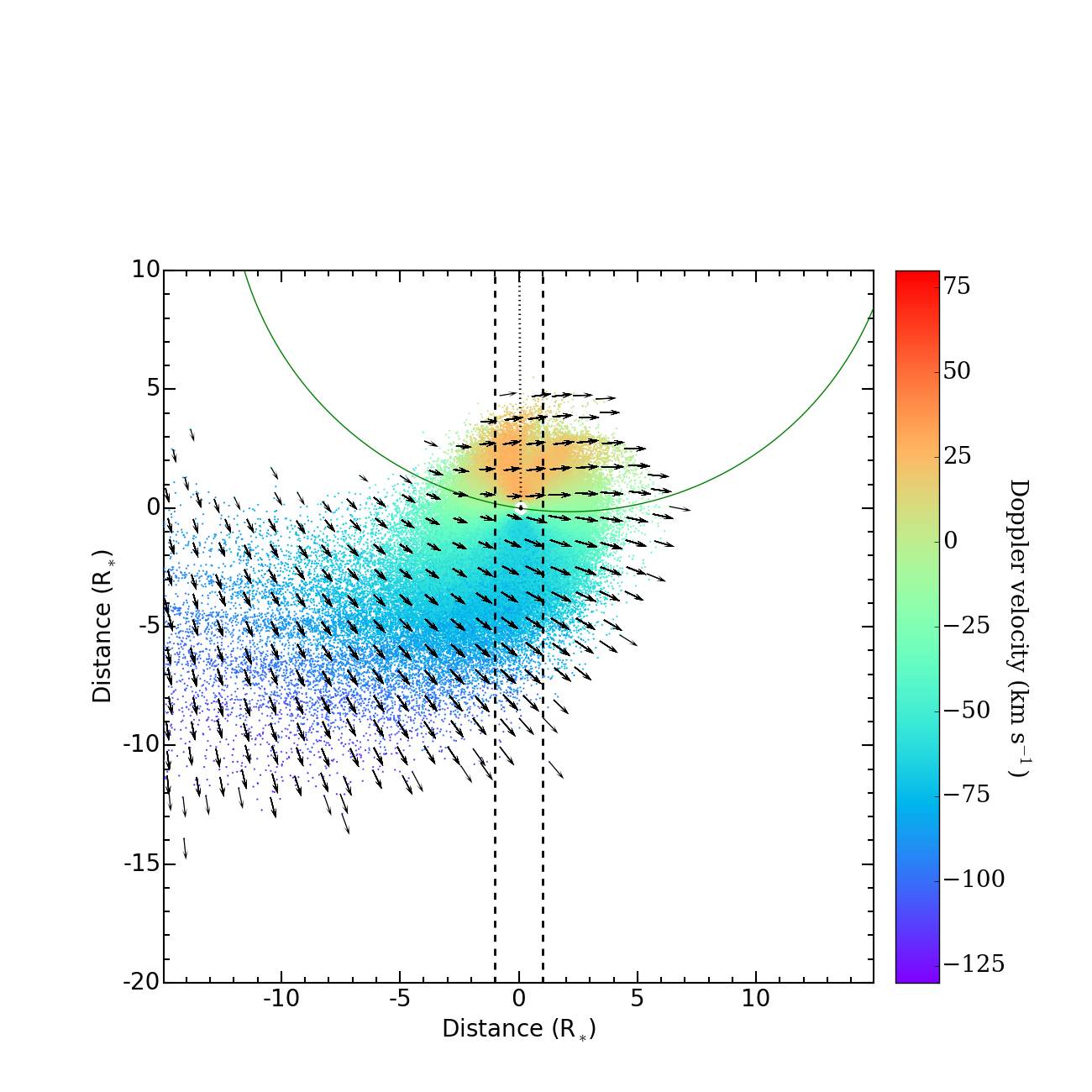}
\includegraphics[trim=4.85cm 1.5cm 1.5cm 7.85cm,clip=true,width=0.356\textwidth]{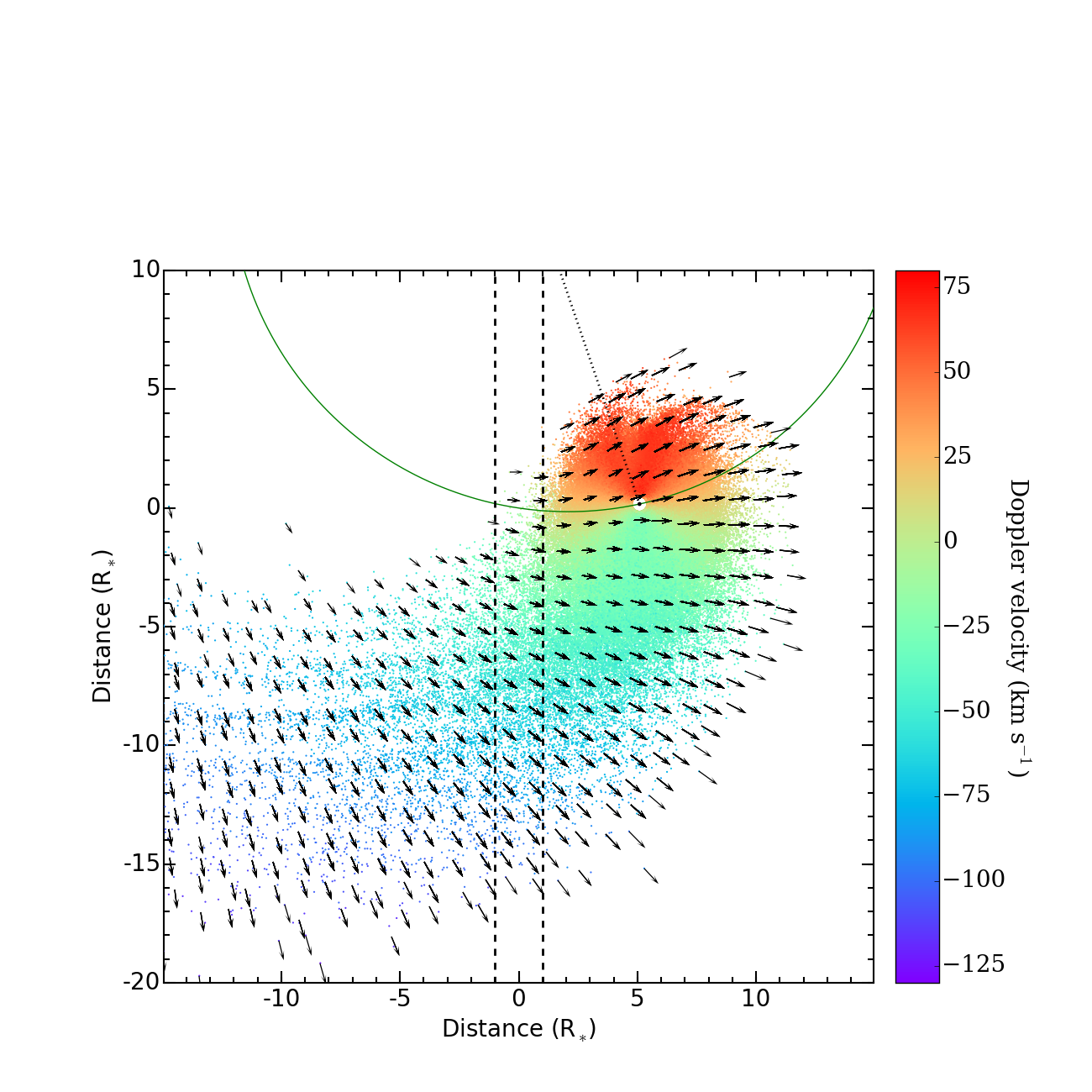}\\
\end{minipage}
\caption[]{Views of GJ\,436b exosphere (lower panels) and its theoretical absorption profile, as it would be observed close to the star (upper panels). We caution that absorption in the range [-40 ; 20]km\,s$^{-1}$ cannot be observed from Earth, and that STIS LSF would spread this absorption signal over several pixels. Time increases from left to right and is indicated on each plot. The dashed black lines limit the LOS toward the stellar surface, while the dotted line indicates the star-planet axis. Black arrows show the velocity field of the H$^{0}$ atoms relative to the star. Gas is seen in the orbital plane and colored as a function of its velocity projected on the LOS. \textit{Left panel:} As they circumvent the planet to move away from the star, atoms in the front of the coma absorb the stellar flux at blueshifted velocities before the optical transit. \textit{Middle panel:} At the time of the optical transit the strongest absorption comes from the core of the coma, and is centered on the projection of the planet orbital velocity at -16\,km\,s$^{-1}$. The outer regions of the exospheric tail contribute to absorption at higher velocities in the blue wing. \textit{Right panel:} After the optical transit, absorption in the blue wing of the line comes exclusively from the tail.}
\label{dyn_time}
\end{figure*}

%%%%%%%%%%%%%%%%%%%%%%%%%%%%%%%%%%%%%%%%%%%%%%%%%%%%%%%%%%%%%%%%%%%%%%%%%%%%%%%%%%%%%%%%%%%%%%%%%%%%%%%%

\subsection{Influence of the physical parameters}
\label{sec:param_influence}

We used simulations performed with EVE to study the effects of the most influential star and planet properties on the structure of the exosphere and its absorption signature. The planetary escape rate and the stellar photo-ionization rate are two basic properties of neutral hydrogen escape, but in contrast to hot-Jupiters subjected to strong radiation pressure the planetary wind velocity also has a significant influence on the exosphere of a moderately irradiated, lower-mass planet like GJ\,436 b.\\

\begin{itemize}
\item \textit{Planetary wind velocity $v_{\mathrm{pl-wind}}$}: This parameter influences the spatial and spectral dispersion of the gas in the exosphere. Higher values for $v_{\mathrm{pl-wind}}$ increase the width of the coma and the exospheric tail, extending in particular the duration of the pre-transit and post-transit absorptions. The same amount of gas escaping faster from the atmosphere is more diffused and has a wider distribution of Doppler velocities, yielding lower depths in the core of the absorption profile but spreading the absorption over a larger spectral range.\\

%----------------------------------------------------
	    
	\item \textit{Photo-ionization rate $\Gamma_{\mathrm{ion}}$:} This parameter acts on the lifetime of neutral hydrogen atoms, and yields local variations in the size of the exosphere and its density. Because of self-shielding, photo-ionization has indeed more impact on the regions facing the star, which leads to a more significant erosion of the tail and the front of the exosphere than its denser coma. Higher values for $\Gamma_{\mathrm{ion}}$ thus decrease more strongly the pre-transit and post-transit absorption depths.\\
   
%----------------------------------------------------
	\item \textit{Escape rate $\dot{M}_{\mathrm{H^{0}}}$:} This parameter changes the local density roughly uniformly within the exosphere. An increase in the escape rate can thus partly balance a stronger photo-ionization as long as self-shielding is not significant. Higher values for $\dot{M}_{\mathrm{H^{0}}}$ increase the depth of the absorption profile at all wavelengths, and lengthen the exospheric tail.\\
	
\end{itemize}
%----------------------------------------------------
 
When hydrogen densities are high enough, self-shielding can protect the interior of the coma from radiation pressure. Gas in these regions and escaping from the planet dayside is subjected to the nominal stellar gravity, and can move far ahead of the planet toward the star. We note however that the very high escape rate/very low photo-ionization rate that would lead to the formation of this extended leading tail are excluded, as it would occult the star too early compared to the observations.

%%%%%%%%%%%%%%%%%%%%%%%%%%%%%%%%%%%%%%%%%%%%%%%%%%%%%%%%%%%%%%%%%%%%%%%%%%%%%%%%%

\section{Interpretation of the observations}
\label{constraints}

\subsection{Qualitative description}
\label{sec:quali}

\begin{figure}    
\includegraphics[trim=0cm 2.2cm 0.5cm 0cm,clip=true,width=\columnwidth]{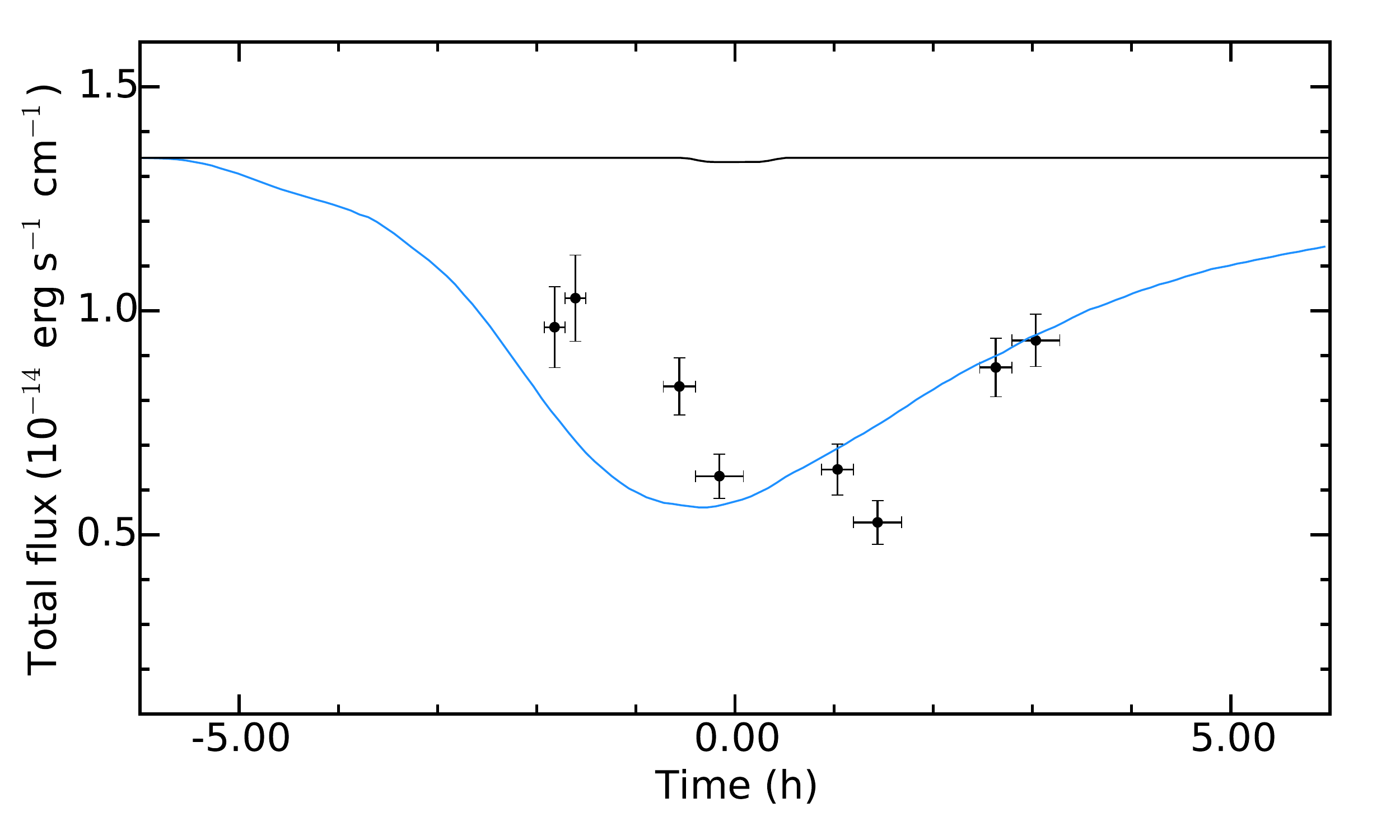}
\includegraphics[trim=0cm 2.2cm 0.5cm 0.5cm,clip=true,width=\columnwidth]{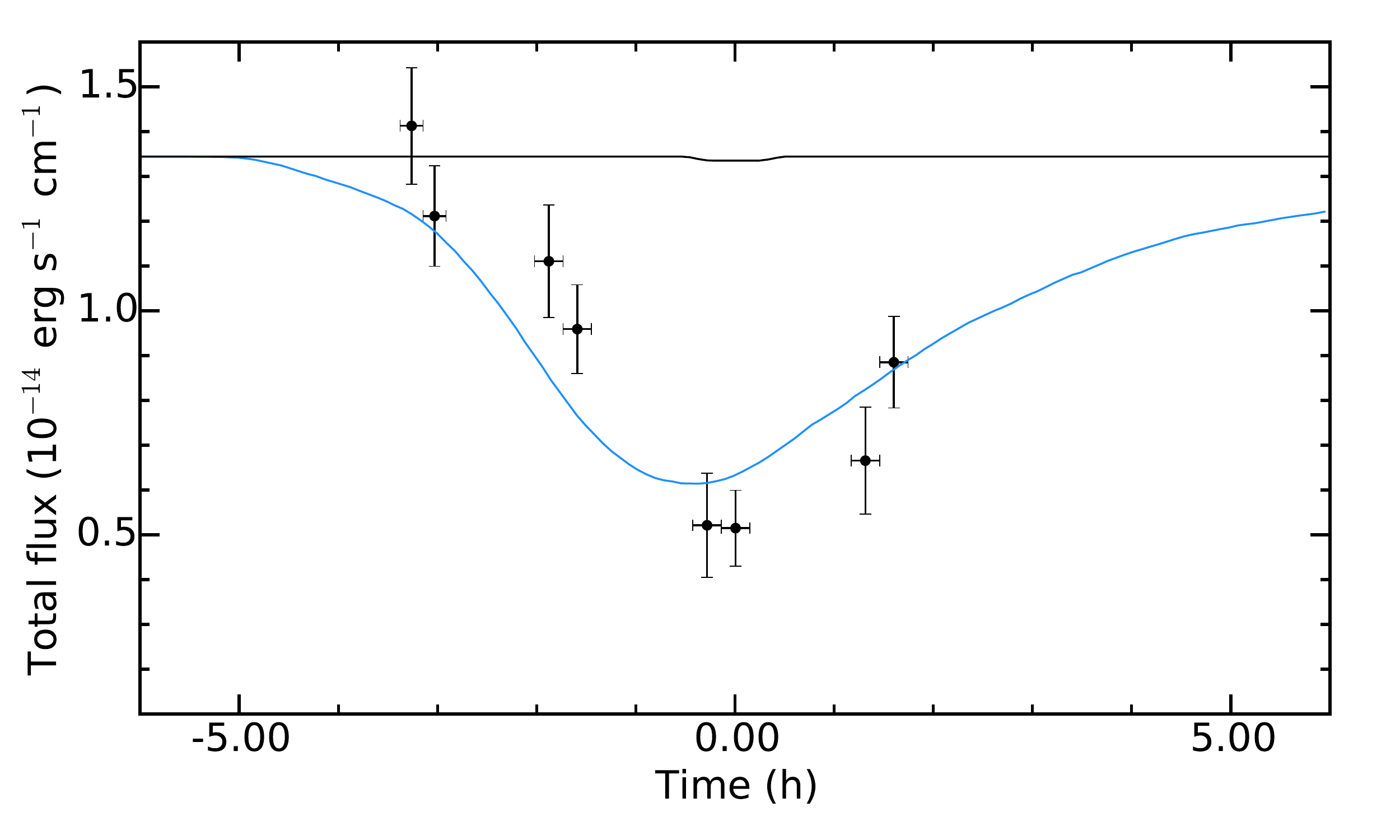}
\includegraphics[trim=0cm 0cm 0.5cm 0.5cm,clip=true,width=\columnwidth]{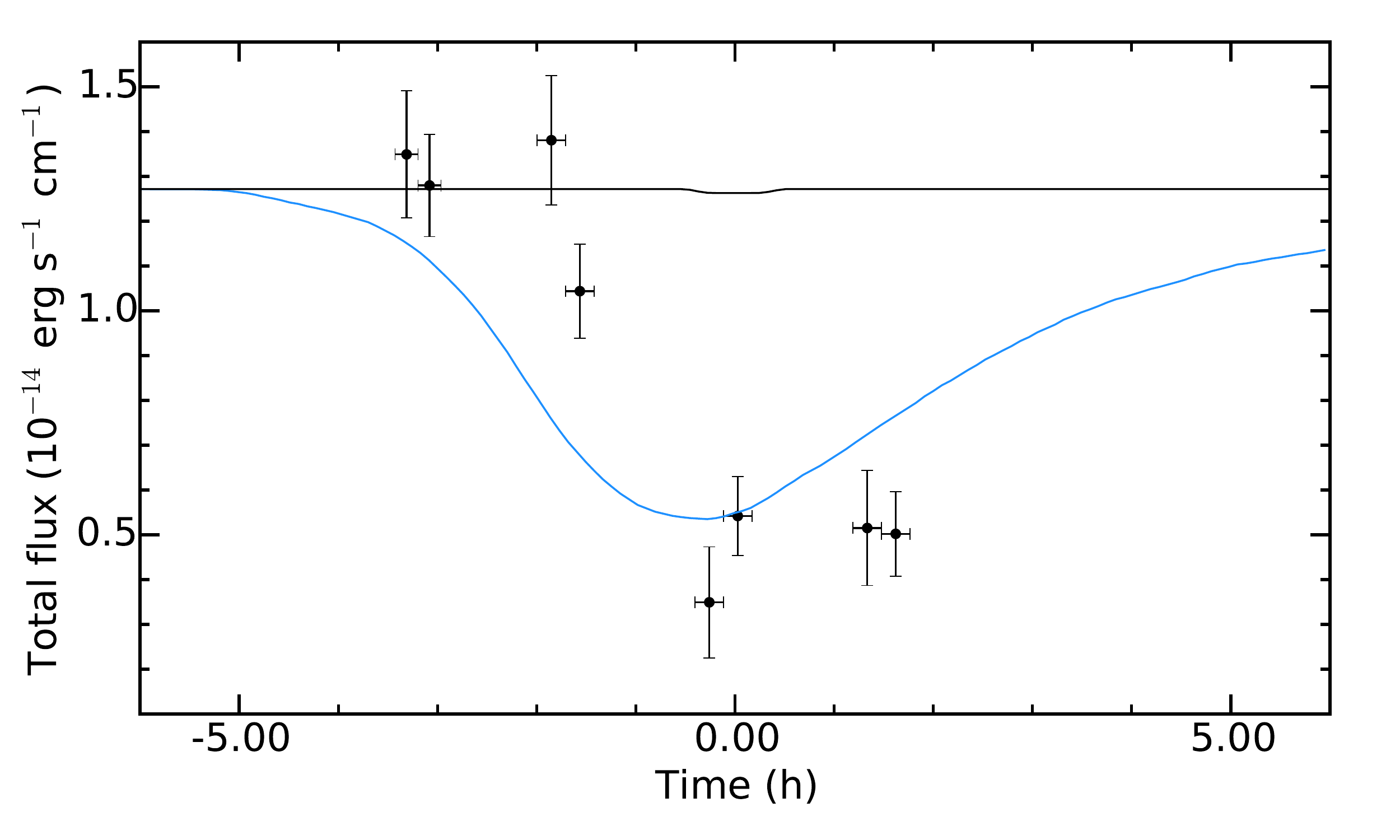}
\caption[]{Transit light curves of the Lyman-$\alpha$ line in Visit 1 (upper panel), Visit 2 (middle panel), and Visit 3 (lower panel), integrated between -120 and -40\,km\,s$^{-1}$. Black points correspond to the observations, while the black line shows the optical transit. The blue lines corresponds to best-fit simulations when the exosphere is subjected to stellar gravity and radiation pressure alone.}
\label{lc_rad_brak}
\end{figure}

The main features of the Lyman-$\alpha$ line observations of GJ\,436 b were described by \citet{Ehrenreich2015}. They showed that the flux in the entire red wing, and at high velocities in the blue wing, is stable over the transit duration and between the different epochs. Visit 1 displays a puzzling absorption in the red wing just after the optical transit, which is not reproduced in subsequent visits. Absorption attributed to the exosphere occurs for all visits in a similar range of velocities up to about -120\,km\,s$^{-1}$ in the blue wing of the line. Overall the light curves measured in this range (Fig.~\ref{lc_rad_brak}) present similar variations over time for the three visits: 1) the exospheric transit begins about 2\,h before the optical transit; 2) the maximum absorption depth is found at about the time of the optical transit; 3) although it decreases after the optical transit, absorption may still be observed for several hours afterwards. There are nevertheless some differences between the visits in the way the absorption varies over time. Visit 1 displays smooth variations during the atmospheric ingress and egress. The decrease in flux is sharper at the ingress of the other visits, in particular in Visit 3. Regarding the post-transit phases, absorption decreases very quickly in Visit 2 while Visit 3 displays absorption depths similar to those of the optical transit. \\

%%%%%%%%%%%%%%%%%%%%%%%%%%%%%%%%%%%%%%%%%%%%%%%%%%%%%%%%

\subsection{Possible scenarios}
\label{sec:scenar}

In this section we investigate different physical scenarios that could explain the features described in Sect.~\ref{sec:quali}.\\

\subsubsection{Radiative braking}
\label{sec:rad_brak_scen}

As shown in Sect.~\ref{dyn_low_prad}, the velocities expected from the dynamics of the radiatively braked exosphere are consistent with the range of the absorption signature up to -120\,km\,s$^{-1}$. Using the model described in Sect.~\ref{model}, we further found that the combination of low radiation pressure with high planetary wind velocities of about 30\,km\,s$^{-1}$ allows the diffusion of the escaping gas within a coma large enough to explain the deep absorption observed in all visits. This wind velocity, higher than those obtained from the modeling of hot-Jupiters expanding thermosphere ($\sim$1-10\,km\,s$^{-1}$; see, e.g.,; \citealt{MurrayClay2009}; \citealt{Koskinen2013a}), may be linked to the lower density of the Neptune-mass planet GJ\,436 b, or caused by a magnetically driven planetary wind (\citealt{Tanaka2014}).\\
The extended coma also produces a deeper and earlier ingress than observed, in particular for Visit 3 (Fig.~\ref{lc_rad_brak}). We also found that the photo-ionization of the gas and its natural dilution as it moves away from the planet leads to a stronger decrease of the absorption depth than observed during the post-transit phases in Visit 3. Therefore, while the exosphere is always subjected to radiation pressure and this force naturally reproduces the observed velocities of the escaping gas, radiative braking alone does not explain the time variations of the measured absorption depth. In other words, radiation pressure is the main driver of the velocity field in the exosphere, and secondary mechanisms must reshape the geometry of the gas cloud.\\

\subsubsection{Radiative blow-out}

As is the case for hot Jupiters, a higher radiation pressure could repel the escaping hydrogen and reduce the front of the coma responsible for an extended early ingress. However, because radiation pressure is proportional to the stellar flux, this would require a higher flux in the entire Lyman-$\alpha$ line, implying that the spectra measured during the first orbits in Visits 2 and 3 are already absorbed by the exosphere. This is unlikely, because these first orbit spectra are remarkably similar (as would be the case if they corresponded to the stable intrinsic stellar Lyman-$\alpha$ line), in contrast to other spectra measured at similar phases of Visits 2 and 3 (Fig.~\ref{lc_rad_brak}). Furthermore to explain absorption as early as the first orbits in Visits 2 and 3, the coma should be even more extended than is the case with a low radiation pressure, in contradiction with its narrowing from the stronger radiation pressure. \\

\subsubsection{Transition region}

\citet{Ehrenreich2015} used the EVE code with similar settings as those described in Sect.~\ref{model} to fit the three combined visits of GJ\,436 b in the frame of a radiative braking scenario. They reproduced the observed sharp early ingress by introducing an additional free parameter to the model, which is a multiplicative factor to artificially increase the ionization cross-section of H$^{0}$ atoms. The corresponding increase in self-shielding coupled with high ionization rates allows for the thin outer envelop of the exosphere facing the star to be eroded, while preserving the denser inner regions. The sharp density transition region thus created is akin to the boundary of a H\,{\sc ii} region, or to a bow-shock environment. However this model requires extremely high ionization rates in the order of 10$^{-6}$\,s$^{-1}$, and makes strong assumptions on self-shielding that affect the overall structure of the exosphere. Detailed models of magnetic interactions between the star and planet will be required to investigate this scenario further. \\

\subsubsection{Stellar wind interactions}
    
Charge exchange between the escaping H$^{0}$ atoms and stellar protons could also ionize the front layers of the coma (e.g., \citealt{Holmstrom2008}) and reduce the depth of the early ingress. These interactions would also ionize planetary neutral atoms in every region of the exosphere, decreasing the overall absorption, but this might be compensated for by the concomitant creation of an independent population of neutralized protons repopulating the exosphere. Moving with the high radial velocity of the stellar wind, this population would be less sensitive to photo-ionization and spatial dilution than the planetary atoms and would keep the same distribution over longer time scales, possibly explaining the post-transit absorption depths in Visit 3 (Sect.~\ref{sec:quali}). A combination of the stable radiation pressure with a variable stellar wind may explain the different variations in absorption depths observed between the three visits.

%%%%%%%%%%%%%%%%%%%%%%%%%%%%%%%%%%%%%%%%%%%%%%%%%%%%%%%%%%%%%%%%%%%%%%%%%%%%%%%%%%%%

\section{Discussion}
\label{conclu}

New HST observations of the Neptune-mass exoplanet GJ 436b revealed a spectacular atmospheric escape (\citealt{Kulow2014}, \citealt{Ehrenreich2015}), repeatable over the three epochs of observations. Where the exospheres of known evaporating hot Jupiters cover $\sim$15\% of the stellar disk, the cloud of hydrogen surrounding GJ\,436 occults up to $\sim$60\% of the star with significant absorption observed several hours before and after the optical transit. The amplitude of the absorption and the corresponding size of the exosphere may appear surprising, considering that GJ\,436b is only moderately irradiated by its M-dwarf host star. \\
To understand this puzzle, we have adapted and run EVE, our 3D particle simulation of atmospheric escape, with the specific properties of the GJ\,436 system. In this paper we focused on a scenario where escaping atoms are submitted to radiation pressure, stellar and planet gravitational forces, and photo-ionization to assess the impact of these phenomenon on the observed Lyman-$\alpha$ transit signature.\\
Radiation pressure is a velocity-dependent force that can be calculated directly from observations of the Lyman-$\alpha$ line, and we provide detailed reconstructions of the intrinsic stellar line at the different epochs. We found that radiation pressure from GJ436 is too low to exceed stellar gravity and to repel the escaping hydrogen atoms away from the star, as is the case for hot Jupiters. Even this moderate radiation pressure is nevertheless enough to brake the gravitational deviation of the gas toward the star, and it allows the formation of a large coma comoving with the planet (although not gravitationally bound to it) that extends in a broad cometary tail trailing the planet and progressively eroded by photo-ionization. \\
Our simulations show that the low radiation pressure from GJ\,436 explains the velocity range of the observed absorption signatures well up to about -120\,km\,s$^{-1}$ and suggest that a high planetary wind velocity of about 30\,km\,s$^{-1}$ is required to create a sizeable coma reproducing the observed absorption depths. However, we found that the variations in the depth and duration of the absorption at the different phases of the transit, some specific to the visit, cannot be explained by radiation pressure alone. In the low radiation pressure field of the star, the coma also extends far ahead of the planet, producing a deeper and earlier ingress than observed. An additional mechanism, such as interactions with the stellar wind, is thus needed to reduce the size of the coma front. By creating a new population of neutral hydrogen atoms with the properties of the stellar wind, charge exchange may also explain peculiar features observed in the last visit.\\
We emphasize that future studies intent on reproducing the observations of GJ\,436 b should take the entire spectral information into account and not only the photometric light curves. Given the size of the exosphere, the regions that were observed transiting the star only sample a small part of the atmospheric structure, and more observations will be required to fully constrain the dynamics of the escaping gas and the properties of GJ\,436 b environment.\\
HST observations of a partially transiting exosphere around the warm giant \mbox{55 Cancri b} (\citealt{Ehrenreich2012}) previously hinted that moderately irradiated planets could have significantly extended exospheres. The detection of GJ 436b exosphere offers the opportunity to probe the atmospheric escape of a ``cool" exoplanet in detail ($\sim$800 K), whose atmospheric photochemistry may also be significantly affected by Lyman-$\alpha$ radiation (\citealt{Miguel2015}). Generalizing these results to other systems, in particular those that will be detected around M dwarfs by CHEOPS, TESS, and PLATO, will open tantalizing new perspectives for characterizating warm and temperate exoplanets.\\

%%%%%%%%%%%%%%%%%%%%%%%%%%%%%%%%%%%%%%%%%%%%%%%%%%%%%%%%%%%%%%%%%%%%%%%%%%%%%%%%%%%%%%%%%%%%%%%%%%%%%%%%%%%%%%%%%%%%%%%%%%

\begin{acknowledgements}
We thank the referee, Jeffrey Linsky, for useful and fruitful comments. This work is based on observations made with the NASA/ESA Hubble Space Telescope, obtained at the Space Telescope Science Institute, which is operated by the Association of Universities for Research in Astronomy, Inc., under NASA contract NAS 5-26555. This work bas been carried out in the framework of the National Centre for Competence in Research ``PlanetS'' supported by the Swiss National Science Foundation (SNSF). V.B. and D.E. acknowledge the financial support of the SNSF. A.L.E acknowledges financial support from the Centre National d\textsc{\char13}Etudes Spatiales (CNES). The authors acknowledge the support of the French Agence Nationale de la Recherche (ANR), under program ANR-12-BS05-0012 ``Exo-Atmos''.  We warmly thank J.B. Delisle and A. Wyttenbach for fruitful discussions. 
\end{acknowledgements}

\bibliographystyle{aa} % style aa.bst
\bibliography{biblio} % your references Yourfile.bib

\end{document}